\pdfoutput=1	% for arXiv % don't have this in the first line when submitting!

\documentclass[final,5p,times,twocolumn]{elsarticle}

\usepackage{graphicx}% Include figure files
\usepackage{amsmath}
\usepackage{color}
\usepackage{bm}	% $\bm{...}$ gives bold maths

\newcounter{bla}

\journal{Computer Physics Communications}

\newcommand{\rmi}{\mathrm{i}}
\newcommand{\rmd}{\mathrm{d}}

\begin{document}

\begin{frontmatter}



\title{Dr TIM: Ray-tracer TIM, with additional specialist scientific capabilities}

\author[Stephen]{Stephen Oxburgh}
\author[Tom]{Tom\'{a}\v{s} Tyc}
\author[Johannes]{Johannes Courtial\corref{author}}
\cortext[author]{Corresponding author\\\textit{E-mail address:} johannes.courtial@glasgow.ac.uk}
\address[Stephen, Johannes]{SUPA, School of Physics \& Astronomy, University of Glasgow, Glasgow G12~8QQ, United~Kingdom}
\address[Tom]{Institute of Theoretical Physics and Astrophysics, Masaryk University, Kotlarska 2,
61137 Brno, Czech~Republic}

\begin{abstract}
We describe several extensions to TIM, a raytracing program for ray-optics research.
These include relativistic raytracing; simulation of the external appearance of Eaton lenses, Luneburg lenses and generalized focusing gradient-index lens (GGRIN) lenses, which are types of perfect imaging devices; raytracing through interfaces between spaces with different optical metrics; and refraction with generalised confocal lenslet arrays, which are particularly versatile METATOYs.
\end{abstract}


\end{frontmatter}

%\submitto{}
%
%
%\pacs{
%01.50.Wg, % (Physics of toys)
%42.15.-i, % (Geometrical optics)
%42.15.Dp, % (Wave fronts and ray tracing)
%42.25.Gy, % (Edge and boundary effects; reflection and refraction)
%42.70.-a% (Optical materials)
%}

\noindent
\textbf{PROGRAM SUMMARY}

\begin{small}
\noindent
{\em Manuscript Title:} Dr TIM: Ray-tracer TIM, with additional specialist scientific capabilities \\
{\em Authors:} Stephen Oxburgh, Tom\'{a}\v{s} Tyc, Johannes Courtial \\
{\em Program Title:} TIM \\
{\em Journal Reference:}                                      \\
  %Leave blank, supplied by Elsevier.
{\em Catalogue identifier:}                                   \\
  %Leave blank, supplied by Elsevier.
{\em Licensing provisions:} GNU GPL \\
  %enter "none" if CPC non-profit use license is sufficient.
{\em Programming language:} Java \\
{\em Computer:} Any computer capable of running the Java Virtual Machine (JVM) 1.6 \\
  %Computer(s) for which program has been designed.
{\em Operating system:}  Any; developed under Mac OS X Version 10.6 and 10.8.3 \\
  %Operating system(s) for which program has been designed.
{\em RAM:} typically 130 MB (interactive version running under Mac OS X Version 10.8.3) \\
  %RAM in bytes required to execute program with typical data.
  % Please give some freely chosen keywords that we can use in a
  % cumulative keyword index.
{\em Classification:} 14 Graphics, 18 Optics \\
  %Classify using CPC Program Library Subject Index, see (
  % http://cpc.cs.qub.ac.uk/subjectIndex/SUBJECT_index.html)
  %e.g. 4.4 Feynman diagrams, 5 Computer Algebra.
{\em Does the new version supersede the previous version?}  yes \\
{\em Reasons for the new version:}
significant extension of capabilities (see \textit{Summary of revisions}), as demanded by our research \\
{\em Summary of revisions:}
added capabilities include 
simulation of different types of camera moving at relativistic speeds relative to the scene;
visualisation of the external appearance of generalized focusing gradient-index (GGRIN) lenses, including Maxwell fisheye, Eaton and Luneburg lenses;
calculation of refraction at the interface between spaces with different optical metrics;
and handling of generalised confocal lenslet arrays (gCLAs), a new type of METATOY \\
{\em External routines/libraries:} JAMA \cite{JAMA-in-summary} (source code included) \\
  % Fill in if necessary, otherwise leave out.
{\em Nature of problem:}
  %Describe the nature of the problem here.
visualisation of scenes that include scene objects that create wave-optically forbidden light-ray fields \\
{\em Solution method:} ray tracing \\
{\em Unusual features:}
  %Describe any unusual features of the program/problem here.
specifically designed to visualise wave-optically forbidden light-ray fields;
can visualise ray trajectories and geometric optic transformations;
can simulate photos taken with different types of camera moving at relativistic speeds,
interfaces between spaces with different optical metrics,
the view through METATOYs and generalised focusing gradient-index lenses;
can create anaglyphs (for viewing with coloured ``3D glasses''), HDMI-1.4a standard 3D images, and random-dot autostereograms of the scene;
integrable into web pages \\
{\em Running time:}\\
  %Give an indication of the typical running time here.
Problem-dependent; typically seconds for a simple scene \\

\end{small}

% \noindent \textit{No.\ of lines in distributed program, including test data, etc.:}  approx.\ 30,900 (incl.\ comments); the source code of JAMA is another approx.\ 3,300 lines

\section{Introduction}

\noindent
TIM \cite{Lambert-et-al-2012} is a ray tracer we developed as a tool for ray-optics research, specifically our work on ``\underline{meta}ma\underline{t}erials f\underline{o}r ra\underline{y}s'' (METATOYs), which are surfaces covered with miniaturised optical components/instruments such that the surface appears to change the direction of transmitted light rays according to very unusual generalised laws of refraction \cite{Hamilton-Courtial-2009}.

The images TIM creates try to be somewhat photo-realistic, but perfect photo-realism is less important to us than simplicity.
Additionally, there is often no obvious unique way in which light behaves at many of TIM's optical elements.
For example, most of TIM's generalised laws of refraction cannot be performed for at least some incident light-ray fields without compromising the integrity of the phase fronts in some way \cite{Courtial-Tyc-2012}, usually by introducing arrays of optical vortices into the field \cite{Courtial-Tyc-2011}.
The optical vortices could, in principle, be added at different positions in the field, and this lack of a unique realisation of these generalised laws of refraction means that there is no obvious way to derive the equivalent of Fresnel coefficients for these laws of refraction.

In addition to being a research tool, TIM is also an outreach tool:
TIM can be compiled into an interactive Java applet (which can be embedded into web pages\footnote{See, for example, \url{http://tinyurl.com/timray}.}) or Java application (which, after downloading, can be run on almost any computer system).
Much of TIM's functionality, including all the extensions described here, can be accessed interactively.
We have tried to design TIM such that it is fun to use and invites playful exploration.

We keep adding capabilities to TIM as demanded by our current research interests, or occasionally just because we think that a feature would increase TIM's appeal as an outreach tool.
Here we describe a few of the extensions that we added since we first wrote about TIM \cite{Lambert-et-al-2012}.
Working in a University, it is only natural to us to recognise TIM's new specialist scientific capabilities by giving the new version the full title ``Dr TIM''.
% \footnote{Please note that the addition to TIM's name is merely an indication of TIM's new specialist scientific capabilities, not an academic title officially conferred by any University.}.

This paper is structured as follows.
In section \ref{relativity-section} we describe TIM's capabilities to model the effect of the scene moving relative to the camera with relativistic speed.
Section \ref{Eaton-et-al-section} describes TIM's simplified raytracing through Maxwell-fisheye, Eaton, Luneburg and generalized focusing gradient-index lens (GGRIN) lenses.
Section \ref{metric-interface-section} outlines TIM's raytracing through the interfaces between spaces with different optical metrics.
In section \ref{gCLAs-section} we derive a law of refraction for generalised confocal lenslet arrays (gCLAs), a METATOY that is becoming increasingly important in our research.
Finally, in section \ref{odds-and-sods-section}, we discuss a number of minor extensions to TIM before we conclude (section \ref{conclusions-section}).
In a number of places throughout the paper we outline details, aimed at readers with some familiarity of the Java programming language, of the way particular tasks have been implemented in the Java code.

\section{\label{relativity-section}Relativistic ray tracing}

\noindent
There are a number of computer programs that visualise the effect of moving at relativistic speed (see, for example, Refs \cite{Howard-et-al-1995,Savage-et-al-2007,Mueller-et-al-2010}).
One of these, \emph{Real Time Relativity} \cite{Savage-et-al-2007}, stands out as it allows the user to move, with relativistic speeds and interactively (in a game-like environment), through complex scenes, and as it can be freely downloaded and run on the most common computer systems.

We have added to TIM the capability to simulate a snapshot taken with a camera that is moving, with relativistic speed, through a scene of stationary objects.
The moment in simulated time when the snapshot is taken can be varied. %; frames of the same scene calculated for different times 
This new capability can be combined with several of TIM's other capabilities, such as simulating a camera with a finite-size aperture and creating anaglyphs for viewing with red-blue 3D goggles.

We simulate a camera moving, with constant velocity $\bm{v}$, through a scene of stationary objects.
We only consider the change in the position in which objects are seen, an effect known as relativistic aberration; we neglect all other effects, specifically the Doppler effect (which alters colour non-isotropically) and the headlight effect (which alters the brightness non-isotropically)~\cite{Savage-et-al-2007}.

\subsection{Calculation of relativistic aberration}

\noindent
We calculate the relativistic aberration by broadly following the approach taken in Ref.\ \cite{Howard-et-al-1995}.

When a camera takes a photo, it briefly opens its shutter, allowing light rays from the scene to enter the camera body and to hit the detector.
Raytracing traces these rays backwards, starting from the camera and into the scene.
If the camera moves relative to the scene, then each backwards-traced light ray needs to be transformed correspondingly when it leaves the camera and enters the scene.
Therefore there are two relevant frames of reference:
the \emph{camera frame}, in which the camera is at rest, and the \emph{scene frame} in which everything else is at rest.
In the scene frame, the camera frame is moving with velocity $\bm{v}$.
We denote space-time coordinates in the camera frame as unprimed, those in the scene frame as primed.
At time $t = 0$, the origins of both frames coincide.
The simplest way to understand the transformation of light rays between the frames is to consider, for each light ray, two events consisting of the light ray passing through two positions on its trajectory, and to transform these between the different frames.
As both frames are inertial, light rays travel in straight lines through those positions in both frames.

Because we are tracing rays backwards, from the camera into the scene, we start in the camera frame.
We do not actually explicitly transform two events in the ray's life.
Instead, we transform one event, and we transform the direction of the backwards-traced ray (which points in the opposite direction of the physical light-ray direction).

In the camera frame, the event happens at position $\bm{x}$ and at time $t$.
If the camera is moving with velocity $\bm{v}$ in the scene frame, we can calculate the event's space-time coordinates in the scene frame by applying the Lorentz transformation \cite{Howard-et-al-1995}
\begin{equation}
\bm{x}^\prime = \bm{x} + (\gamma - 1) \frac{(\bm{\beta} \cdot \bm{x}) \bm{\beta}}{\beta^2} + \gamma \bm{\beta} c t, \quad
c t^\prime = \gamma c t + \gamma (\bm{\beta} \cdot \bm{x}), \label{Lorentz-transformation}
\end{equation}
where $\bm{\beta} = \bm{v}/c$ and $\gamma = 1/\sqrt{1-\beta^2}$ is the Lorentz factor.
Note that we do not need to calculate the time of the event in the scene frame, as we assume that the scene is not changing.
We can transform the direction $\bm{d}$ of the backwards-traced ray by considering two events, namely the positions $\bm{x}_1$ and $\bm{x}_2 = \bm{x}_1 - \bm{d}$ on its trajectory and the times $t_1$ and $t_2 = t_1 + \|d\| / c$ the ray passes through these (note the different signs in the expressions for $\bm{x}_2$ and $t_2$).
Lorentz-transforming the positions of these events (according to Eqn (\ref{Lorentz-transformation})) gives
\begin{align}
\bm{x}_1^\prime
&= \bm{x}_1 + (\gamma - 1) \frac{(\bm{\beta} \cdot \bm{x}_1) \bm{\beta}}{\beta^2} + \gamma \bm{\beta} c t_1, \\
\bm{x}_2^\prime
&= \bm{x}_2 + (\gamma - 1) \frac{(\bm{\beta} \cdot \bm{x}_2) \bm{\beta}}{\beta^2} + \gamma \bm{\beta} c t_2
%&= \bm{x}_1 - \bm{d} + (\gamma - 1) \frac{(\bm{\beta} \cdot (\bm{x}_1 - \bm{d})) \bm{\beta}}{\beta^2} + \gamma \bm{\beta} c (t_1 + \|d\|/c) \\
%&= \bm{x}_1 + (\gamma - 1) \frac{(\bm{\beta} \cdot \bm{x}_1) \bm{\beta}}{\beta^2} + \gamma \bm{\beta} c t_1 \nonumber \\
%&- \bm{d} - (\gamma - 1) \frac{(\bm{\beta} \cdot \bm{d}) \bm{\beta}}{\beta^2} + \gamma \bm{\beta} \|d\| \\
= \bm{x}_1^\prime - \bm{d}^\prime,
\end{align}
where
\begin{align}
\bm{d}^\prime
= \bm{d} + (\gamma - 1) \frac{(\bm{\beta} \cdot \bm{d}) \bm{\beta}}{\beta^2} - \gamma \bm{\beta} \|d\|
\end{align}
is the direction of the backwards-traced ray in the scene frame.

%\item The light ray that eventually hits a specific pixel passes through the position conjugate to the pixel position, i.e.\ the position where the camera's lens (or more general focussing element --- note that TIM can focus on non-planar surfaces \cite{Lambert-et-al-2012}) produces the image of the pixel.

%But as the scene frame is an inertial frame, light rays travel in straight lines in this frame and so we now know all about the light ray in the scene frame we need to know:
%it hits the position $\bm{x}_1^\prime$ with direction $\bm{d} = \bm{x}_2^\prime - \bm{x}_1^\prime$.

We choose as the event the light ray leaving the last object it encounters that is stationary in the camera frame.
We call the collection of these objects the \emph{camera-frame scene}.
TIM's backwards raytracing proceeds by each ray leaving the entrance-pupil disk and
interacting with any objects in the camera-frame scene, before being transformed into the scene frame and interacting with any objects in the scene.
The underlying assumption is that the objects in the camera frame are closer to the camera than any objects in the scene frame, and so the rendering only works correctly if the camera-frame scene is populated accordingly and the shutter timing is selected appropriately (as the shutter time determines the position of the shutter surface at the time the shutter opens).

\begin{figure}
\begin{center}
\includegraphics{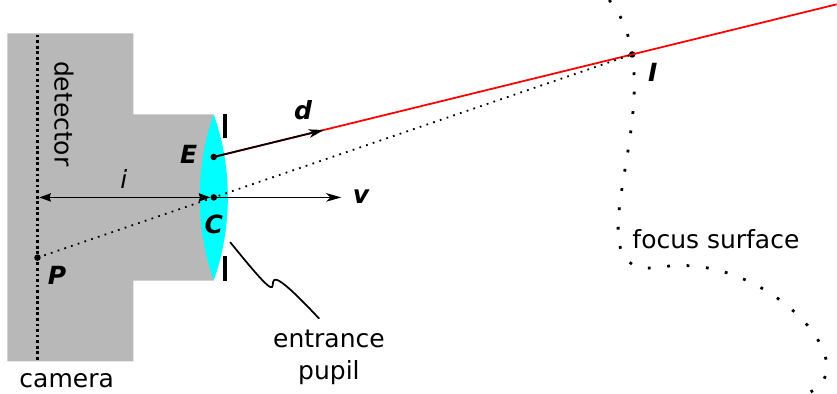}
\end{center}
\caption{\label{camera-figure}Camera and shutter model in the camera frame.
For each detector pixel, the camera traces one or more rays from different points on the disk of the entrance pupil to the image of that detector pixel, $\bm{I}$.
If the light ray passes through the entrance-pupil disk at the point $\bm{E}$, then the direction of the backwards-traced ray is $\bm{d} \propto \bm{I} - \bm{E}$.
The camera's entrance-pupil disk is defined by its centre point $\bm{c}$, the view direction $\bm{v}$, which is perpendicular to the disk, and the disk radius (not shown).}
\end{figure}

It is clear from Eqn (\ref{Lorentz-transformation}) that the time of the ray leaving the camera frame has an effect on the position of the same event in the scene frame, and this time is determined by the shutter model used.
Each camera has a shutter somewhere within its body, in modern SLRs usually immediately in front of the film plane (``focal-plane shutter'').
TIM's shutter model allows a choice of three \emph{shutter surfaces} in which the shutter can be placed (Fig.\ \ref{camera-figure}):
\begin{enumerate}
\item the \emph{detector plane}, which is perpendicular to the view direction $\bm{v}$ and positioned a distance $i$ behind the entrance pupil;
\item the \emph{entrance-pupil disk}, positioned immediately in front of the lens (or generalised focussing element; note that TIM can focus on non-planar surfaces \cite{Lambert-et-al-2012});
\item the \emph{focus surface}, the surface on which the images (formed by the focussing element) of the detector pixels are located.
\end{enumerate}
The shutter model assumes that the shutter opens for an instant, at time $t_s$, simultaneously (in the camera frame) across the entire shutter surface.

Irrespective which shutter model is selected, each backwards-traced ray that contributes towards a particular detector pixel originates from a point $\bm{E}$ on the entrance-pupil disk in the direction of the position $\bm{I}$ of the image of the detector pixel (see Fig.\ \ref{camera-figure}).
In the entrance-pupil shutter model, the ray originates from $\bm{E}$ at time $t_s$.
In the focus-surface shutter model, the ray originates from $\bm{E}$ at a time $t_s - \| \bm{I} - \bm{E} \| / c$, so that it passes through the position $\bm{I}$ --- which lies on the focus surface --- at time $t_s$.

The detector-plane shutter model is most complicated.
It makes the following assumptions about the imaging element:
\begin{enumerate}
\item Any light ray through the centre of the entrance pupil, $\bm{C}$, is undeviated.
\item The optical path length from the position of a pixel, $\bm{P}$, to the position of its image, $\bm{I}$, is the same irrespective of the point on the entrance pupil the light ray passes through.
\item The optical path length from $\bm{P}$ to $\bm{I}$ equals $\| \bm{I} - \bm{P} \|$.
\end{enumerate}
The first two assumptions are generalisations of the properties of an idealised lens;
the third assumption is unrealistic, but amongst the available arbitrary choices it is arguably the simplest.
The ray originates from $\bm{E}$ at such a time that it reaches $\bm{I}$ at the same time as the (undeviated) ray that leaves the pixel position $\bm{P}$ at time $t_s$ and that reaches $\bm{I}$ via the aperture centre $\bm{C}$, which reaches $\bm{I}$ at $t_s + \| \bm{I} - \bm{P} \| / c$.
As the ray from $\bm{E}$ takes a time $\| \bm{I} - \bm{E} \| / c$ to travel to $\bm{I}$, it has to leave $\bm{E}$ at time $t_s + (\| \bm{I} - \bm{P} \| - \| \bm{I} - \bm{E} \| ) / c$.
In line with most other implementations of relativistic ray tracing, TIM uses a value for the speed of light of $c = 1$.
%Note that the value of $c$ doesn't matter here.
%It only matters when making a movie!

\subsection{Examples and discussion}

% lattice parameters (standard lattice that appears when initialising scene to "Relativistic")
%   x_min = -1.5, x_max = 1.5, no of cylinders 4
%   y_min = -1.5, y_max = 1.5, no of cylinders 4
%   z_min = -1, z_max = 10, no of cylinders 12

\begin{figure}
\begin{center}
\begin{tabular}{cc}
\raisebox{5.5cm}{(a)} & \includegraphics[width=0.9 \columnwidth]{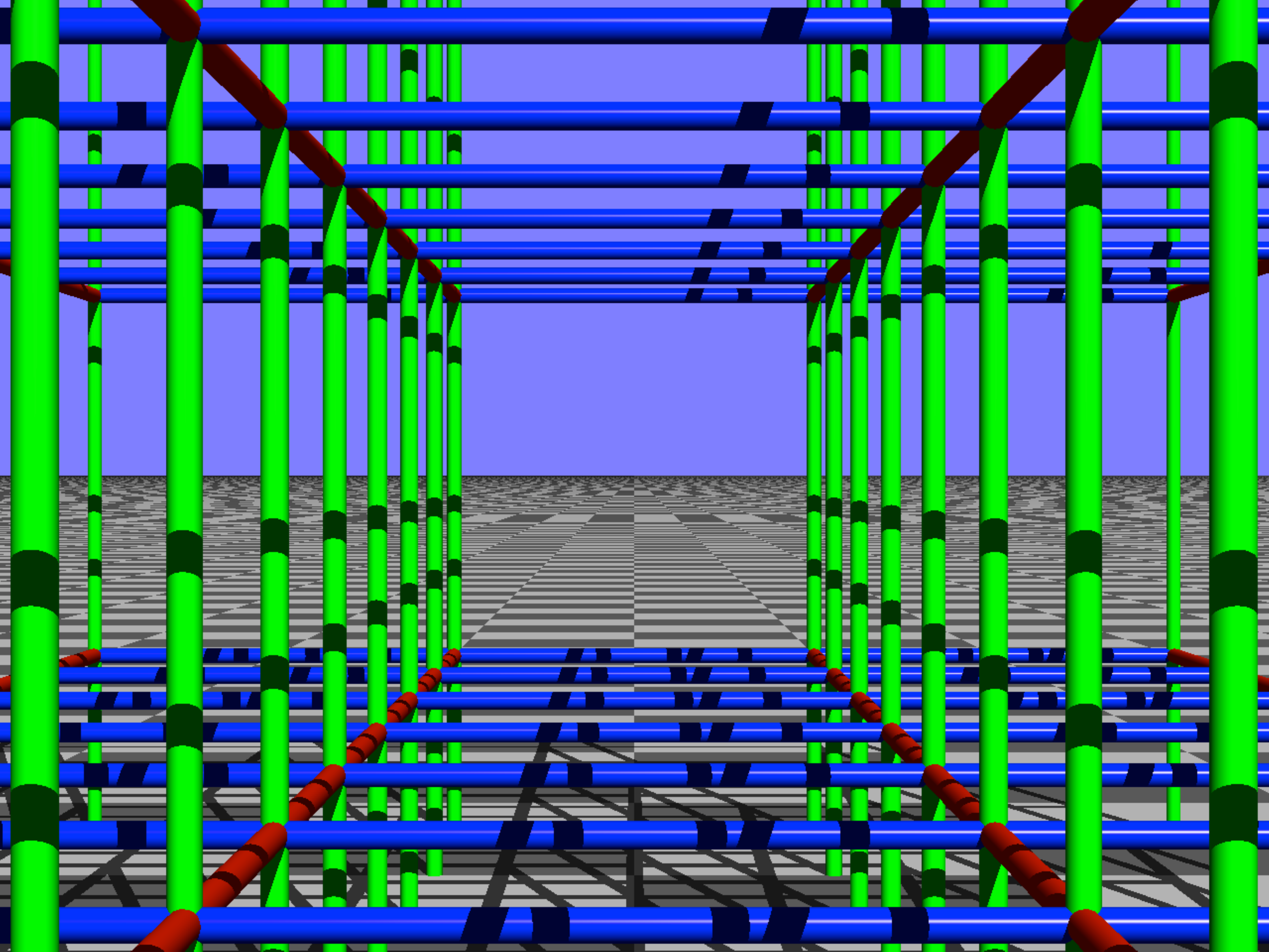} \\
\raisebox{5.5cm}{(b)} & \includegraphics[width=0.9 \columnwidth]{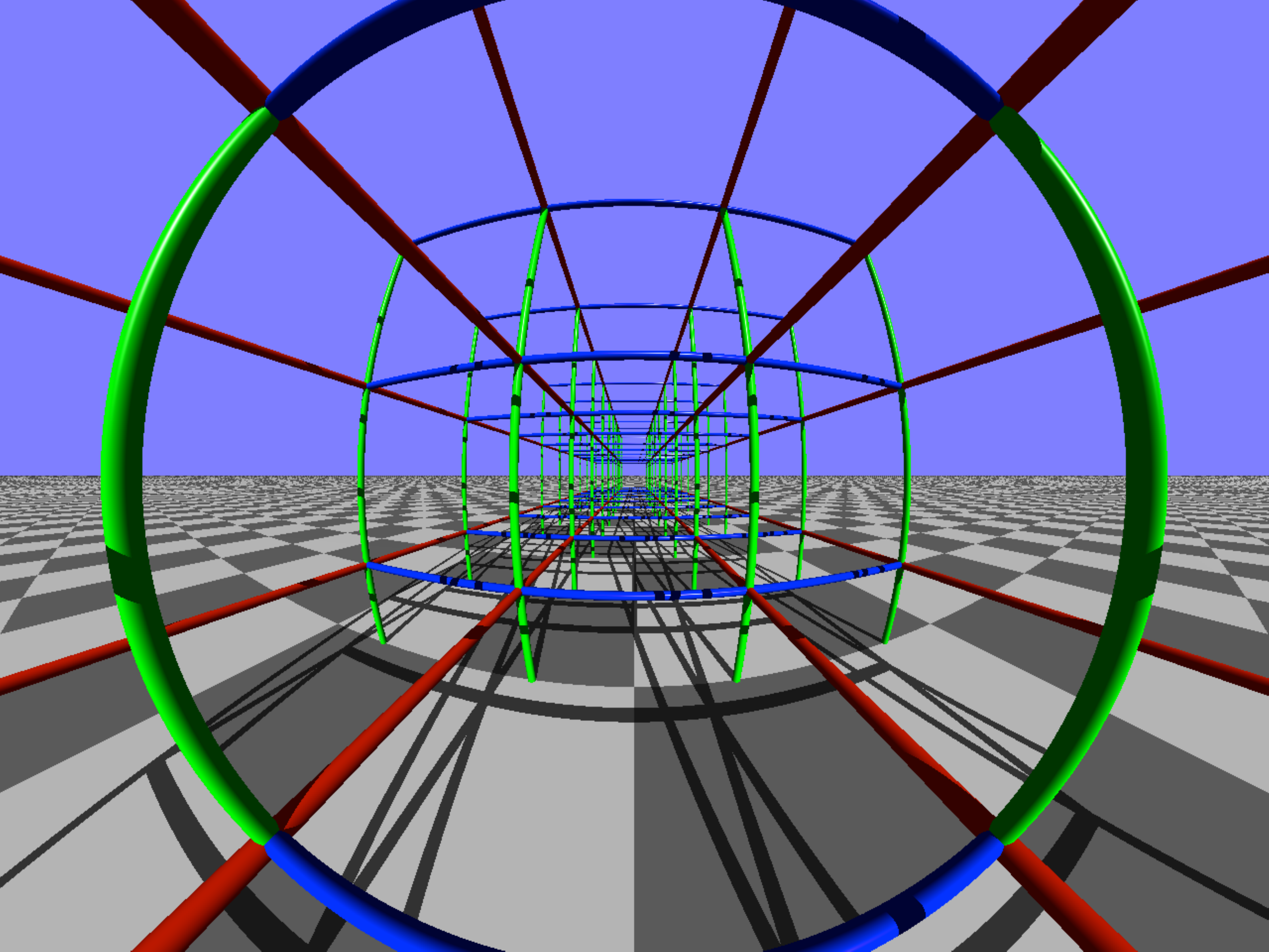} \\
\raisebox{5.5cm}{(c)} & \includegraphics[width=0.9 \columnwidth]{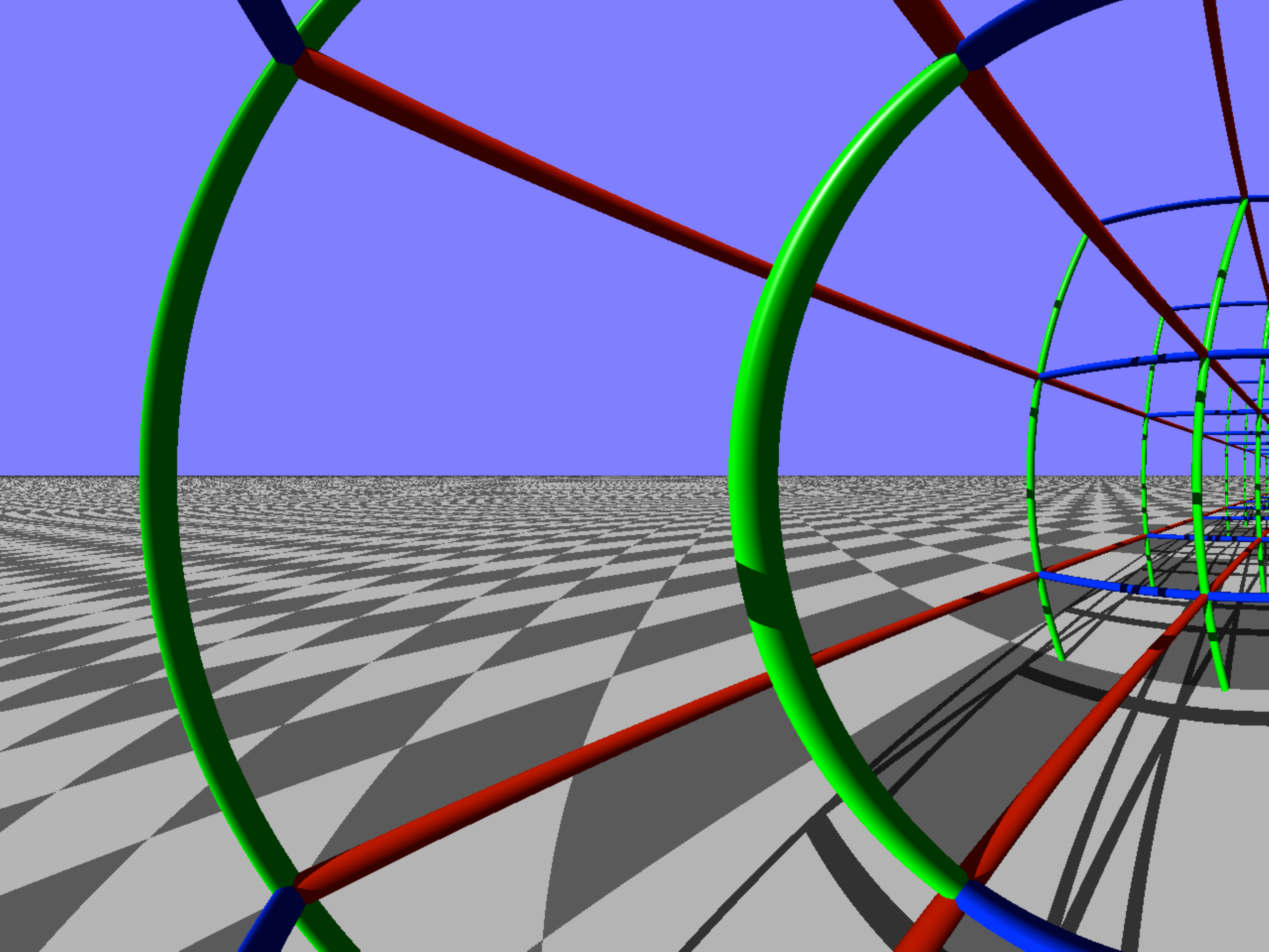}
\end{tabular}
\end{center}
\caption{\label{pinhole-raytracing-figure}Raytracing simulation of a photo of a cylinder lattice taken with a pinhole camera moving with velocity $\bm{v}$ relative to the scene.
(a)~$\bm{v}=(0, 0, 0)$;
(b)~$\bm{v}=(0, 0, 0.99 c)$;
(c)~$\bm{v}=(0.2 c, 0, 0.97 c)$.
The camera is pointing in the $z$ direction.}
% Anti-aliasing quality: Good
\end{figure}

\noindent
Figure \ref{pinhole-raytracing-figure} shows simulated photos taken with a rapidly moving pinhole camera, created with TIM.
A number of well-known effects of relativistic aberration are visible, such as Penrose's result that straight lines  in one frame are seen as straight lines or circular arcs in the other \cite{Penrose-1959}.

\begin{figure}
\begin{center}
\includegraphics[width=\columnwidth]{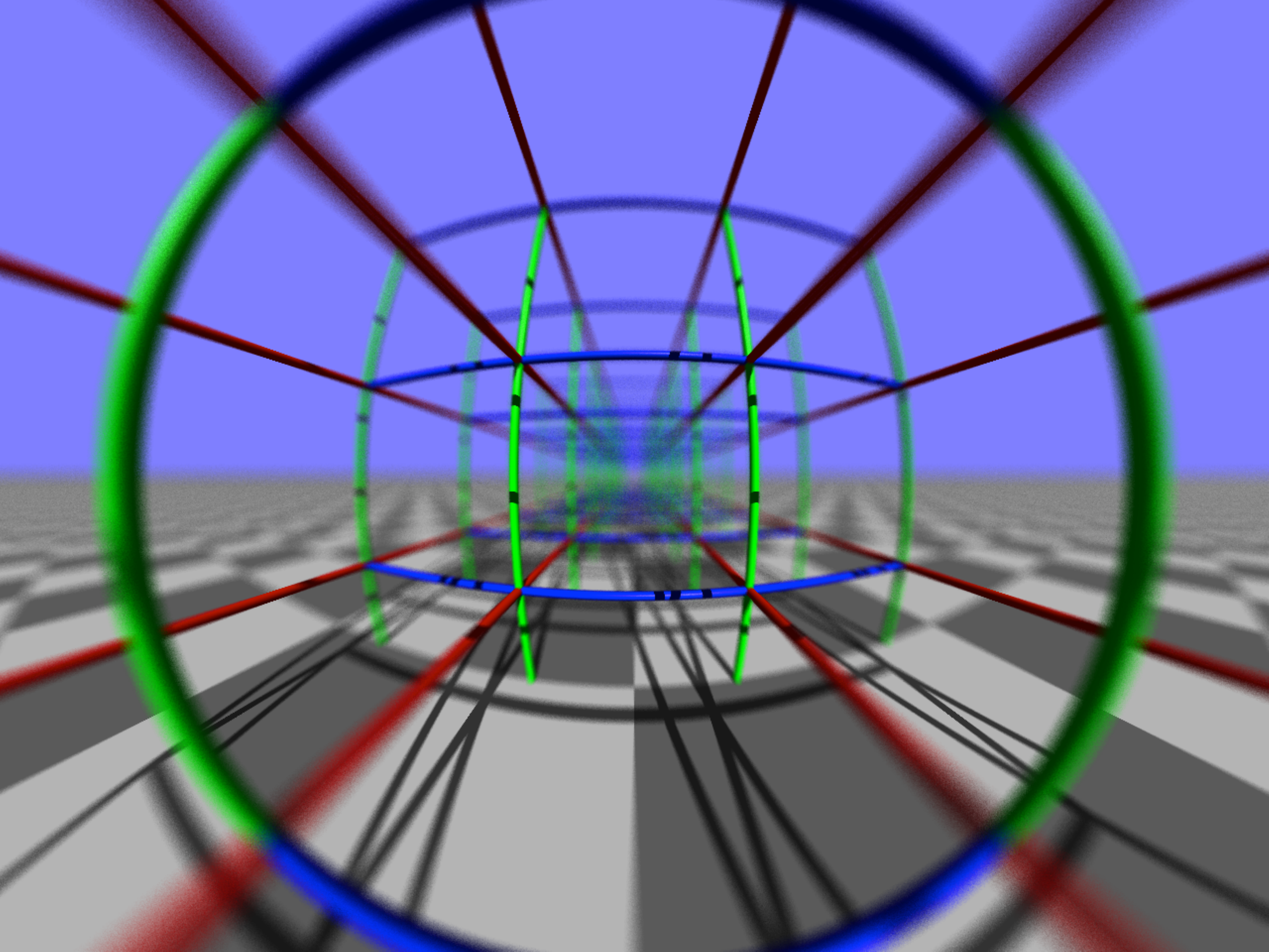}
\end{center}
\caption{\label{blurred-raytracing-figure}
Raytracing simulation of a photo of a cylinder lattice, taken with a camera moving with velocity $\bm{v}=(0, 0, 0.99 c)$ relative to the scene and with a finite-size aperture.
The camera is focussed, in the camera frame, on a plane a distance 15 floor-tile lengths in front of the camera.
The entrance-pupil shutter opened at time $t = 0$.}
% entrance-pupil shutter
% focussing distance: 15
% Aperture size: Medium
% Blur quality: Great
% Anti-aliasing quality: Good
% took about 23 minutes to render
\end{figure}

\begin{figure}
\begin{center}
\includegraphics[width=\columnwidth]{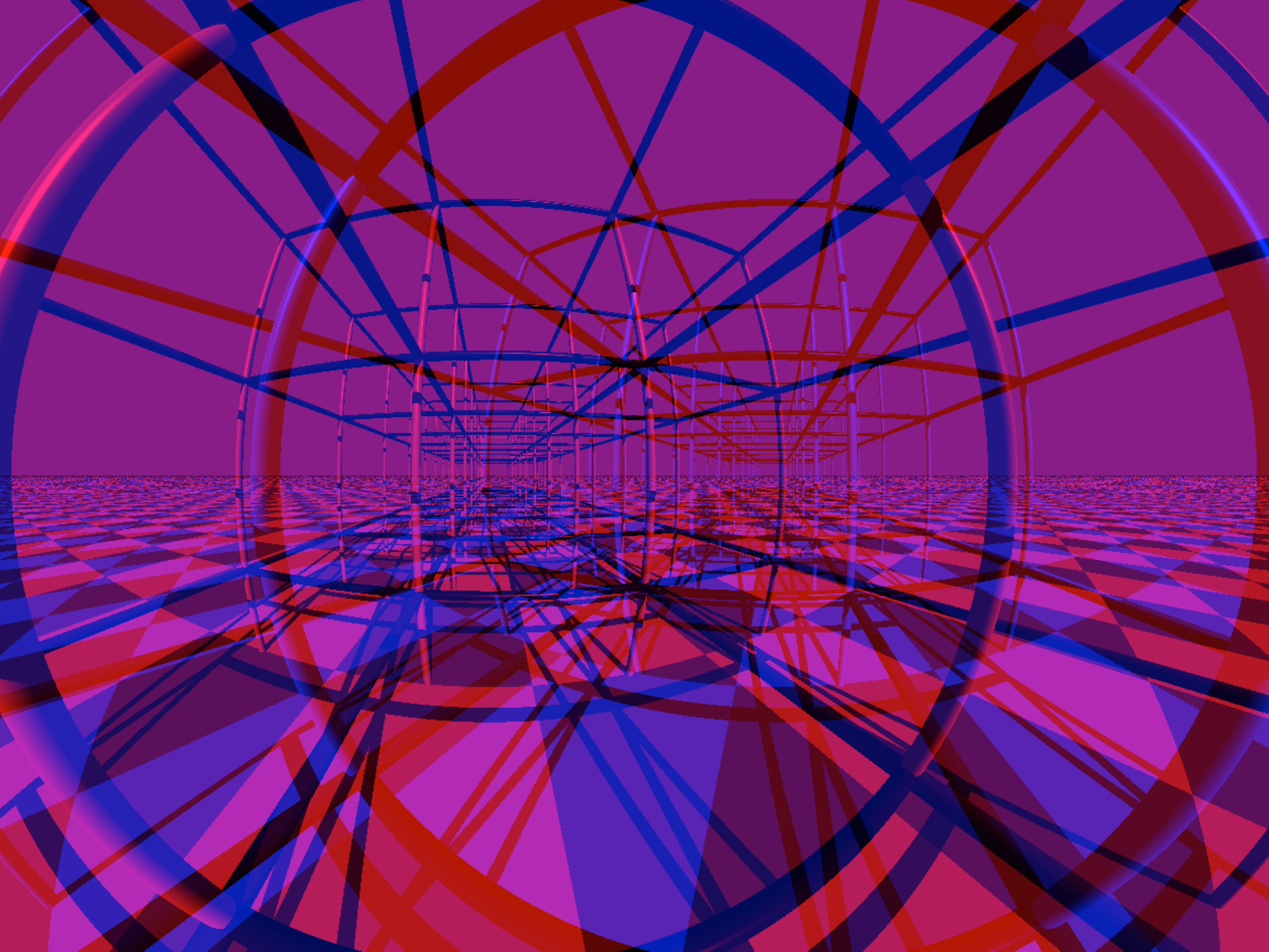}
\end{center}
\caption{\label{anaglyph-raytracing-figure}
Relativistic analyph.
The image shows a raytracing simulation of a photo of a cylinder lattice taken with two cameras moving with velocity $\bm{v}=(0, 0, 0.99 c)$ relative to the scene.
The image recorded by one camera is shown in red, that recorded by the other camera is shown in blue, which makes the image suitable for viewing with red-blue analyph glasses.
The image is simulated for an eye separation is 0.4 floor-tile lengths in the $x$ direction (in the camera frame).}
% centre of view (0, 0, 5)
% eye separation (0.4, 0, 0)
% Anti-aliasing quality: Good
\end{figure}

The calculation of relativistic aberration applies not only to pinhole cameras, but also to other cameras implemented in TIM, and we have extended TIM accordingly.
The resulting combination of capabilities is, to the best of our knowledge, new.
Firstly, TIM can simulate a camera with a finite-size circular aperture.
TIM's simulated camera can focus on a transverse plane in front of the camera, like a standard camera, but it can also focus on almost arbitrary surfaces \cite{Lambert-et-al-2012}.
In the new, relativistic, TIM, this surface is interpreted to be in the camera frame, which is a natural generalisation of the fact that the focussing distance of a standard camera would also refer to the focussing distance in the camera frame.
Figure \ref{blurred-raytracing-figure} shows an example of a simulated photo taken with such a camera.
Secondly, TIM can now create analyphs of the scene as seen by a binocular observer moving with a relativistic speed.
An example is shown in Fig.\ \ref{anaglyph-raytracing-figure}.

% (In focus-surface-shutter mode, the ray's start point gets transformed, and can end up inside an object.  Try standard Tim, beta = (0, 0, 0.99), 

\subsection{Implementation details}

\noindent
The Java classes describing the cameras outlined above are \texttt{RelativisticAnyFocusSurfaceCamera}, which is a subclass of \texttt{AnyFocusSurfaceCamera}, and \texttt{RelativisticAnaglyphCamera}, which is a modification of the (no longer existing) class \texttt{AnaglyphCamera}.
All are part of the package \texttt{optics.raytrace.cameras}.
For use in TIM's interactive version, there are also editable versions of these cameras, called \texttt{EditableRelativisticAnyFocusSurfaceCamera} and \texttt{EditableRelativisticAnaglyphCamera}, which are part of the \texttt{optics.raytrace.GUI.cameras} package.

Lorentz transforms are performed by the new class \texttt{LorentzTransform}, which is part of the \texttt{math} package.

In order to allow backwards raytracing through a camera-frame scene before Lorentz-transforming, it was necessary to keep track of time when tracing light rays.
This required extensive changes throughout the code.
Occasionally, when there was no unique (and easy to implement) way to calculate the time taken to traverse a specific optical component, TIM simply uses the time taken to traverse the equivalent length of vacuum.

%\begin{figure}
%\input{backwardsRays.pdf_tex}
%\caption{\label{backwards-rays-figure}Rays that reach the camera by the time the source of the ray is behind the camera.
%Such rays are not correctly traced in TIM.
%In the example shown here, a light ray leaves from a point $\bm{P}$ on the surface of an object.
%}
%\end{figure}
%
%\emph{Nonsense, I think:}
%A few things are not accurately represented in the model TIM uses.
%It is noticeable that part of the simulated photo, namely anything outside the blue-green circle (which is how the cylinders in the lattice that are in the same plane as the camera show up) shows objects \emph{behind} the camera.
%This makes sense as the light rays would have left the object before they reached the camera.
%But as the camera moves at a very high speed relative to the scene, the object from which such rays originated might well be behind the camera at the time the light ray reaches the camera (see Fig.\ \ref{backwards-rays-figure}).
%\textit{But such light rays can't move through objects that obscure them in any frame.  Is all this chat about seeing the far side of objects moving at relativistic speeds bollocks?}

%\textit{Could easily be generalised to several objects moving at different relativistic speeds in different directions.
%Need to transform each ray from camera into each object's rest frame and calculate which of any intersected objects is being intersected first.
%But have to think about how to handle mirrors etc.\ moving with relativistic speeds; project with Martin?}

\section{\label{Eaton-et-al-section}Maxwell fisheye, Eaton, Luneburg, and generalized focusing 
gradient-index (GGRIN) lenses}

\begin{figure}
\begin{center}
\includegraphics{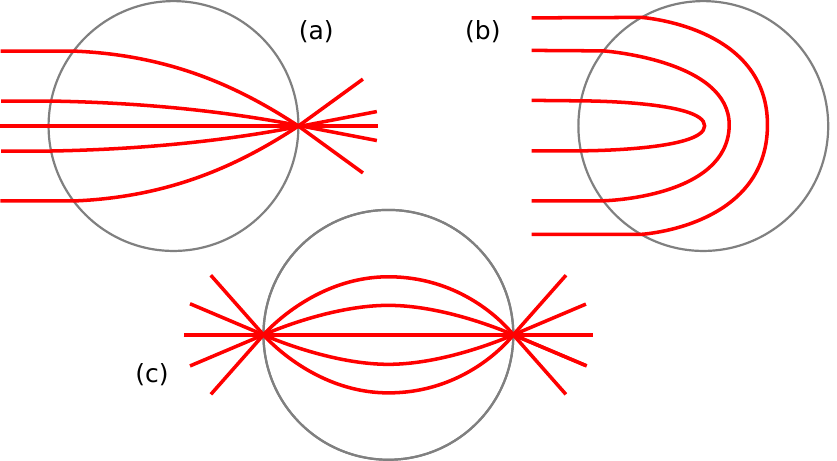}
\end{center}
\caption{\label{lenses-figure}Ray trajectories (red lines) through a Luneburg lens~(a), Eaton lens~(b), and Maxwell-fisheye lens~(c).
The surface of each lens is shown as a grey circle.
(After Ref.\ \cite{Sarbort-Tyc-2012}.)}
\end{figure}

\noindent
Recent work on perfect imaging has revived interest in refractive-index distributions with ray-optically perfect imaging properties.
The Luneburg lens \cite{Luneburg-1944} is a sphere whose inside has a spherically symmetric refractive-index distribution which images any parallel ray bundle incident on the sphere onto a point on the opposite side (Fig. \ref{lenses-figure}(a)).
A slightly different, but still spherically symmetric, refractive-index distribution results in the Eaton lens \cite{Eaton-1952}, which is a perfect retro-reflector (Fig.\ \ref{lenses-figure}(b)).
A Maxwell fisheye \cite{Maxwell-1854} is, in principle, a spherically symmetric refractive-index distribution of infinite radius.
It is famous not only for being the first of these lenses, but also for perfectly imaging not just ray-optically, but also wave-optically \cite{Leonhardt-2009}.
Following Ref.\ \cite{Sarbort-Tyc-2012}, we consider here only a spherical central part of the Maxwell fisheye with its radius chosen such that any point on the outside of the sphere is imaged to the opposite side of the sphere (Fig.\ \ref{lenses-figure}(c)).
We refer to such a device as a Maxwell-fisheye lens.
All of these, and other, ``lenses'',  were recently combined into a more general class by \v{S}arbort and Tyc~\cite{Sarbort-Tyc-2012}.
We call those general lenses \emph{generalized focusing gradient-index (GGRIN) lenses}.

\begin{figure}
\begin{center}
\includegraphics[width=\columnwidth]{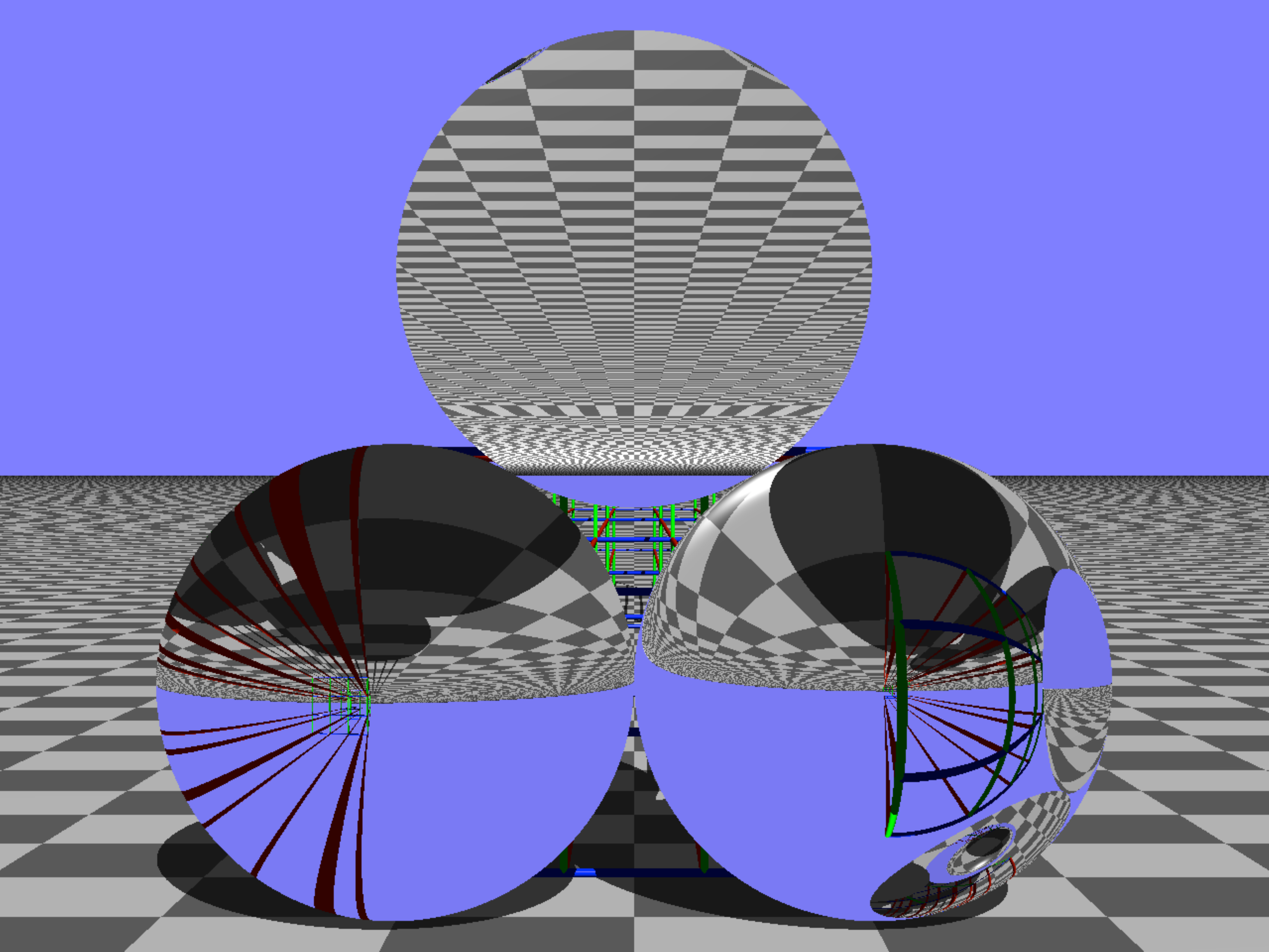}
\end{center}
\caption{\label{lenses-pile-figure}Simulated view of a Luneburg lens (bottom left), Eaton lens (top), and Maxwell-fisheye lens (bottom right) in front of TIM's standard cylinder lattice.
The lattice cannot be seen through the Eaton lens, as it is a retro-reflector.}
% Luneburg lens centred at (-1, 0, 10)
% Eaton lens centred at (0, 1.732, 10)
% Maxwell-fisheye lens centred at (1, 0, 10)
% camera's aperture centre at (0, 0.866 (= 1.732/2), 0)
% Anti-aliasing quality = Good
% Horizontal angle of view = 30¡
\end{figure}

We have extended TIM to be able to simulate the appearance of such lenses (Fig.\ \ref{lenses-pile-figure}).
TIM does not perform detailed raytracing through the lenses.
Instead, TIM uses equations for the position $\bm{P}^\prime$ on the sphere (of radius $R$ and centred at position $\bm{C}$) and direction $\bm{d}^\prime$ after transmission through the lens of a light ray incident at position $\bm{P}$ with direction $\bm{d}$.

\begin{figure}
\begin{center}
\includegraphics{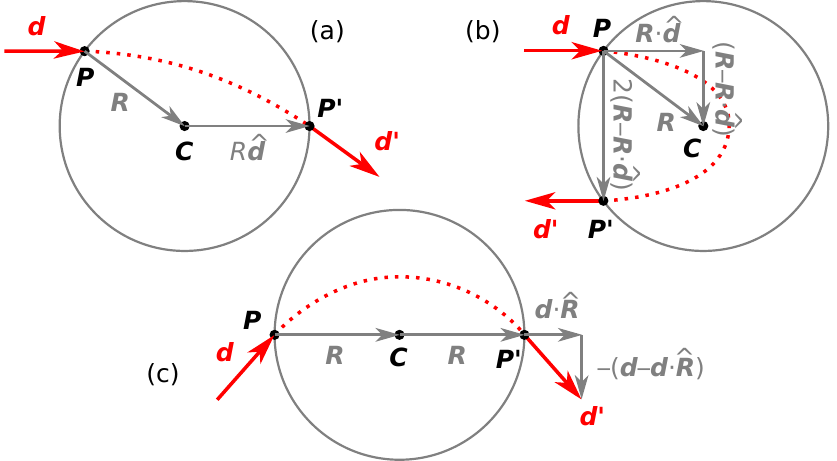}
\end{center}
\caption{\label{diagram}Redirection of a light ray by a Luneburg lens~(a), an Eaton lens~(b), and a Maxwell-fisheye lens~(c).
A light ray hits the lens at $\bm{P}$ with direction $\bm{d}$, and leaves it from position $\bm{P}^\prime$ with direction $\bm{d}^\prime$.}
\end{figure}

For the Luneburg lens, $\bm{P}^\prime$ and $\bm{d}^\prime$ can be calculated as follows.
From Fig.\ \ref{diagram}(a) it can be seen that the outgoing ray leaves the lens from a position
\begin{equation}
\bm{P}^\prime = \bm{C} + R \hat{\bm{d}}
\end{equation}
with in the (unnormalised) direction
\begin{equation}
\bm{d}^\prime = \bm{R} = \bm{C} - \bm{P}.
\end{equation}

For the calculation of the parameters of the ray leaving the Eaton lens, refer to Fig.\ \ref{diagram}(b).
We first calculate the vector $\bm{R} = \bm{C} - \bm{P}$ again, and its component perpendicular to $\bm{d}$, $\bm{R} - \bm{R} \cdot \hat{\bm{d}}$, where $\hat{\bm{d}} = \bm{d} / \| \bm{d} \|$ is the normalised incident light-ray direction.
Then the position $\bm{P}^\prime$ where the outgoing ray leaves the lens is simply
\begin{equation}
\bm{P}^\prime = \bm{P} + 2 \left(\bm{R} - \bm{R} \cdot \hat{\bm{d}} \right),
\end{equation}
and the direction of the outgoing ray is
\begin{equation}
\bm{d}^\prime = - \bm{d}.
\end{equation}

For the Maxwell-fisheye lens (see Fig.\ \ref{diagram}(c)), we calculate the position from which the ray leaves the lens as
\begin{equation}
\bm{P}^\prime = \bm{C} + \bm{R},
\end{equation}
where $\bm{R} = \bm{C} - \bm{P}$ as before, and its direction as the incident light-ray direction, but with the component perpendicular to $\bm{R}$ reversed:
\begin{equation}
\bm{d}^\prime = \bm{d} \cdot \hat{\bm{R}} - \left(\bm{d} - \bm{d} \cdot \hat{\bm{R}} \right)
= 2 \bm{d} \cdot \hat{\bm{R}} - \bm{d}.
\end{equation}

\subsection{GGRIN lens}

\begin{figure}
\begin{center}
\includegraphics{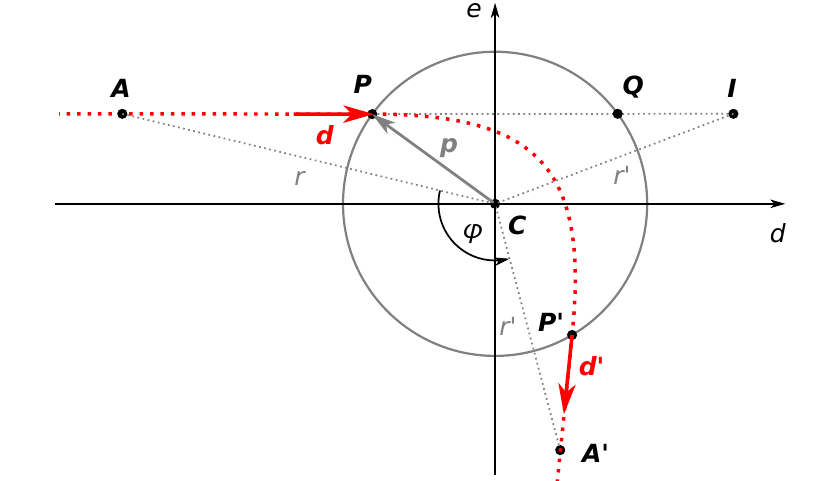}
\end{center}
\caption{\label{Sarbort-Tyc-lens-diagram}Geometry of the effect of the GGRIN lens \cite{Sarbort-Tyc-2012}.
The entire ray trajectory lies within the ray's plane of incidence.}
\end{figure}

\noindent
All the lenses discussed above, and more, can be viewed as special cases of a more general GGRIN lens \cite{Sarbort-Tyc-2012}.
The GGRIN lens is spherically symmetric, and so any ray incident on the lens traverses it along a trajectory that lies in the ray's plane of incidence.
Fig.\ \ref{Sarbort-Tyc-lens-diagram} shows a schematic diagram of an arbitrary ray in its plane of incidence.
If the lens is centred at $\bm{C}$, and the ray hits the lens at position $\bm{P}$ with direction $\bm{d}$, the plane of incidence includes the positions $\bm{C}$ and $\bm{P}$ and the direction vector $\bm{d}$.
The lens can then be characterised by two distances, $r$ and $r^\prime$, and an angle, $\varphi$, as follows.
Before hitting the lens, the (straight-line) ray trajectory passes through a position $\bm{A}$ a distance $r$ from the lens centre, $\bm{C}$.
After transmission through the lens, the ray will pass through a position $\bm{A}^\prime$ a distance $r^\prime$ from $\bm{C}$ such that the distance between the ray's straight-line continuation and the lens centre, $\bm{C}$, remains unchanged (angular-momentum conservation). % impact parameter
$\bm{A}^\prime$ is positioned such that the straight lines $\bm{AC}$ and $\bm{CA}^\prime$ intersect at an angle $\varphi$.
Table \ref{ST-lens-parameters-table} lists the values of the parameters $r$, $r^\prime$ and $\varphi$ for which the GGRIN lens is equivalent to the lenses discussed above.
(Note that the lens with $r = r^\prime = \infty$ and values of $\varphi$ other than $0$ and $180^\circ$ has an effect similar to that of an axicon \cite{McLeod-1954}, which can be used to turn plane waves into Bessel-like light beams~\cite{Herman-Wiggins-1991}.
Unlike axicons, the GGRIN lens should work for plane waves incident from any direction.)
% Is the refractive-index distribution just linearly dependent on r, by any chance?

\begin{table}
\begin{center}
\begin{tabular}{l|ccc}
lens & $r$ & $r^\prime$ & $\varphi$ \\
\hline
Luneburg & $\infty$ & $R$ & $180^\circ$ \\
Eaton & $\infty$ & $\infty$ & $0^\circ$ \\
Maxwell fisheye & $R$ & $R$ & $180^\circ$
% invisible sphere & $\infty $ & $\infty$ & $180^\circ$
\end{tabular}
\end{center}
\caption{\label{ST-lens-parameters-table}Parameter combinations for which a GGRIN lens is equivalent to other well-known lenses.
$R$ is the radius of the lens.}
\end{table}

We calculate the position $\bm{P}^\prime$ where the transmitted ray leaves the lens and its direction $\bm{d}^\prime$ as follows.
First, we define orthogonal unit vectors that span the ray's plane of incidence.
We take the normalised incident light-ray direction, $\hat{\bm{d}} = \bm{d} / \| \bm{d} \|$, as the first unit vector.
Then the component of the vector $\bm{p} = \bm{P} - \bm{C}$ in the direction of $\hat{\bm{d}}$, i.e.\ the $d$ component of $\bm{p}$, is $p_d = \hat{\bm{d}} \cdot \bm{p}$.
We define the second unit vector, $\hat{\bm{e}}$, such that the $e$ component of $\bm{p}$ is positive and $\hat{\bm{d}} \cdot \hat{\bm{e}} = 0$.
The part of $\bm{p}$ in the $e$ direction is $\bm{p} - \hat{\bm{d}} p_d$, and so we define
\begin{equation}
\hat{\bm{e}} = \frac{\bm{p} - \hat{\bm{d}} p_d}{| \bm{p} - \hat{\bm{d}} p_d |}.
\end{equation}
The $e$ component of $\bm{p}$ is then simply $p_e = \hat{\bm{e}} \cdot \bm{p}$.

We can now calculate the $d$ and $e$ components of the vector $\bm{a} = \bm{A} - \bm{C}$.
As $\bm{A}$ lies on the trajectory of the incident ray, $a_e = p_e$.
The $d$ component can be calculated from the distance requirement, i.e.\ $a_d^2 + a_e^2 = r^2$, which gives $a_d = - \sqrt{r^2 - a_e^2}$ (the negative sign is chosen so that $\bm{A}$ actually lies on the incident light-ray trajectory).

We now consider a hypothetical ray that passes through the lens undeviated.
We define $\bm{Q}$ to be the position on the surface of the lens where this ray would re-emerge, and we define $\bm{I}$ to be the position where the ray's distance from $\bm{C}$ is $r^\prime$.
The ray has most of the properties of the refracted ray:
the ray passes through a position, $\bm{I}$, a distance $r^\prime$ from $\bm{C}$, and its impact parameter (the distance between its straight-line continuation and $\bm{C}$) is the same as that of the incident ray (because they share the same straight-line continuation and therefore impact parameter);
all that is wrong is that, in general, the straight lines $\bm{AC}$ and $\bm{CI}$ do not intersect at the desired angle $\varphi$.
We define $\bm{q} = \bm{Q} - \bm{C}$ and $\bm{i} = \bm{I} - \bm{C}$, whose components are
\begin{align}
q_e = i_e = p_e, \quad q_d = - p_d, \quad i_d = \sqrt{r^{\prime 2} - i_e^2}.
\end{align}

The point $\bm{I}$ has all the properties of the point $\bm{A}^\prime$, apart from its direction with respect to $\bm{C}$.
To construct $\bm{A}^\prime$, we need to rotate $\bm{I}$ in the plane around $\bm{C}$ by an angle
\begin{equation}
\beta = \arctan(a_d, a_e) - \arctan(i_d, i_e) + \varphi
\end{equation}
(where we have used the 2-argument arctan function, sometimes called atan2, that avoids quadrant ambiguity).

\begin{figure}
\begin{center}
\includegraphics[width=\columnwidth]{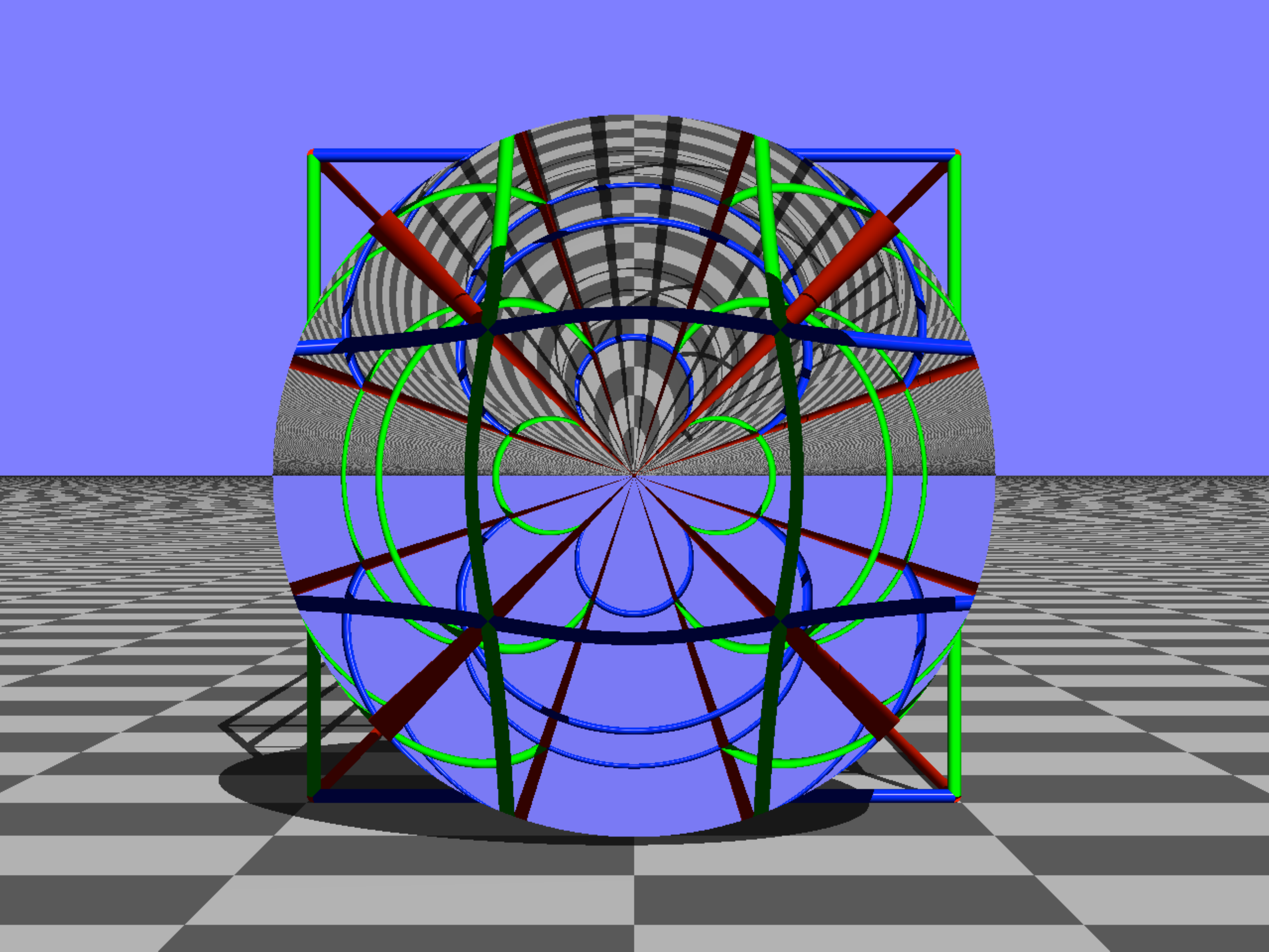}
\end{center}
\caption{\label{ST-lens-view-figure}Simulated view of TIM's standard cylinder lattice seen through an example of a GGRIN lens.
The lens parameters are $r = r^\prime = 100000 \, (\approx \infty)$, $\varphi = 170^\circ$.}
\end{figure}

Similarly rotating $\bm{Q}$ gives the point $\bm{P}^\prime$ where the transmitted ray leaves the lens,  and rotating $\hat{\bm{d}}$ results in the transmitted ray's direction, $\hat{\bm{d}}^\prime$.
We do this by calculating the rotated basis vectors
\begin{align}
\hat{\bm{d}}^\prime = \hat{\bm{d}} \cos \beta + \hat{\bm{e}} \sin \beta, \quad
\hat{\bm{e}}^\prime = -\hat{\bm{d}} \sin \beta + \hat{\bm{e}} \cos \beta,
\end{align}
and calculating the transmitted ray's start point according to
\begin{align}
\bm{P}^\prime = \bm{C} + q_d \hat{\bm{d}}^\prime + q_e \hat{\bm{e}}^\prime;
\end{align}
the transmitted ray's direction is simply $\hat{\bm{d}}^\prime$.
Fig.\ \ref{ST-lens-view-figure} shows an example of the view through a GGRIN lens.

\subsection{Combinations of Eaton and Luneburg lenses with an invisible sphere}

\begin{figure}
\begin{center}
\includegraphics{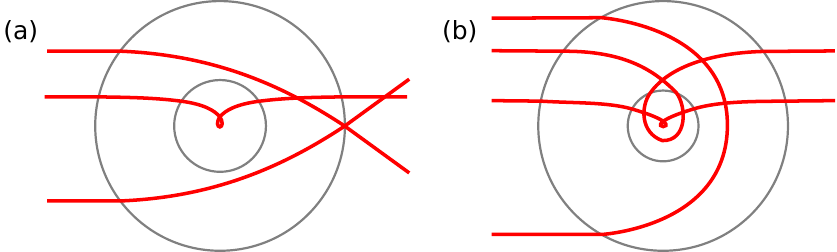}
\end{center}
\caption{\label{lenses-plus-invisible-sphere-figure}Ray trajectories (red lines) through combinations of a Luneburg lens with an invisible sphere~(a) and an Eaton lens with an invisible sphere~(b).
The outer grey circle is the surface of each lens.
The inner grey circle shows the boundary between the Luneburg or Eaton part of the lens (outside) and the invisible-sphere part (inside).
(After Ref.\ \cite{Sarbort-Tyc-2013}.)}
\end{figure}

\noindent
In a further generalisation \cite{Sarbort-Tyc-2013}, \v{S}arbort and Tyc considered multi-focal versions of these lenses, formed by replacing the central part with another refractive-index distribution.
We have also implemented here the ability to render the view through multi-focal lenses formed by combining any of the lenses discussed above with an invisible sphere \cite{Perczel-et-al-2011}.

\begin{figure}
\begin{center}
\begin{tabular}{cc}
\raisebox{4.5cm}{(a)} & \includegraphics[width=0.8 \columnwidth]{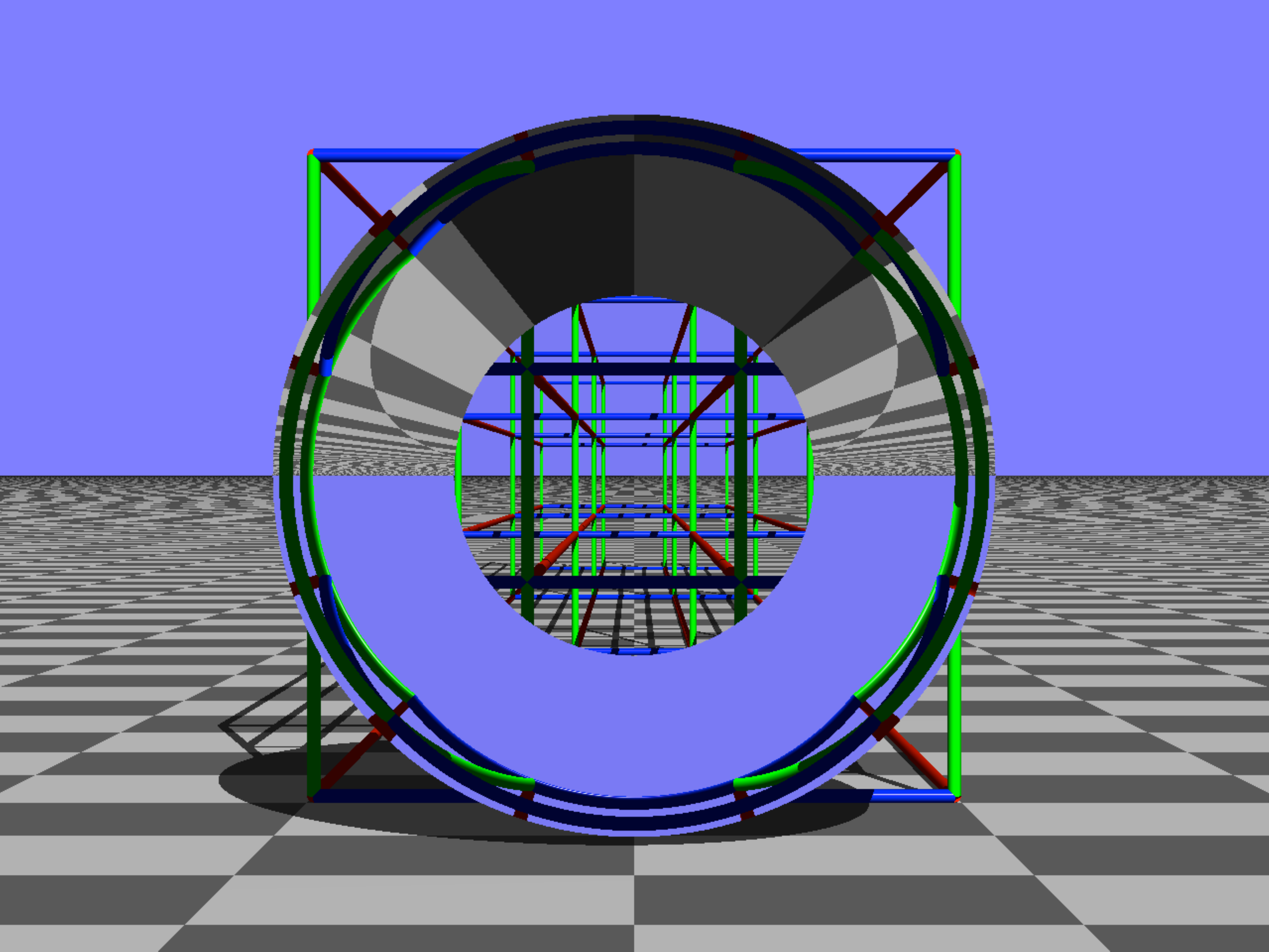} \\
\raisebox{4.5cm}{(b)} & \includegraphics[width=0.8 \columnwidth]{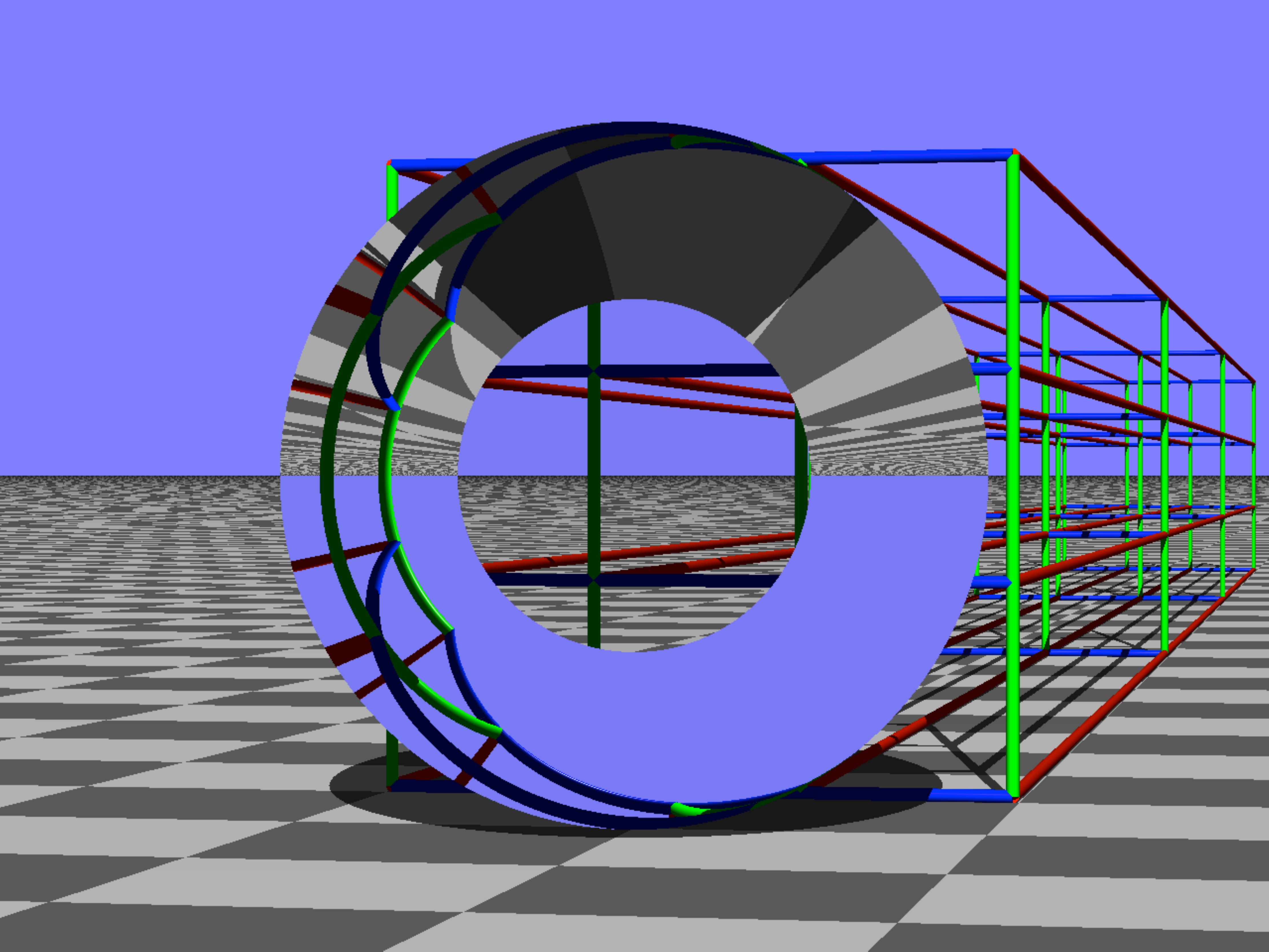}
\end{tabular}
\end{center}
\caption{\label{transparent-tunnel-figure}View of a multifocal lens in front of TIM's standard cylinder lattice, seen from two view positions.
The multifocal lens is a combination of a GGRIN lens (parameters $r = r^\prime = 100000 \, (\approx \infty)$, $\varphi = 90^\circ$) and an invisible sphere.
The appearance is that of a central empty tunnel in the view direction, irrespective of view direction.
(a)~Standard camera position, $(0, 0, 0)$, and view direction, $(0, 0, 1)$;
(b)~view position $(2, 0, 0)$, view direction $(-0.2, 0, 1)$.}
\end{figure}
% ST lens R=1, r=r'=100000, phi=90¡, nRatio=1, R_t=0.5
% side view: aperture centre = (2, 0, 0), view direction = (-0.2, 0, 1)
% front view: aperture centre = (0, 0, 0), view direction = (0, 0, 1) (i.e. standard parameters)

Fig.\ \ref{lenses-plus-invisible-sphere-figure} shows a few ray trajectories through two of these multifocal lenses.
What they have in common is that rays whose straight-line continuation gets closer to the lens centre than $R_t$ emerge from the other side of the lens as if they have passed straight through.
When seen from any direction, the lens appears to have an empty central tunnel of radius $R_t$ in the current view direction.
Fig.\ \ref{transparent-tunnel-figure} shows an example.

\subsection{Refractive-index ratio at the edge}

\noindent
All the above lenses can be designed such that the refractive index on the surface of the lens is the same as that of the surrounding medium.
We have also added to TIM the possibility of simulating the view when this is not the case.

All this requires is a refraction step, from the surrounding medium into a material with the refractive index of the lens at the surface, before the calculations outlined above, and afterwards another refraction step, this time from a material with the lens's surface refractive index into the surrounding medium.
As a respectable ray tracer, TIM is already equipped with refraction functionality, and so this is simply a matter of invoking the relevant method.
Note that the actual value of the refractive indices at the surface and of the surrounding medium do not matter; all that matters is the value of their ratio, and so this is the parameter that can be set for any of the above lenses.

\begin{figure}
\begin{center}
\includegraphics[width=\columnwidth]{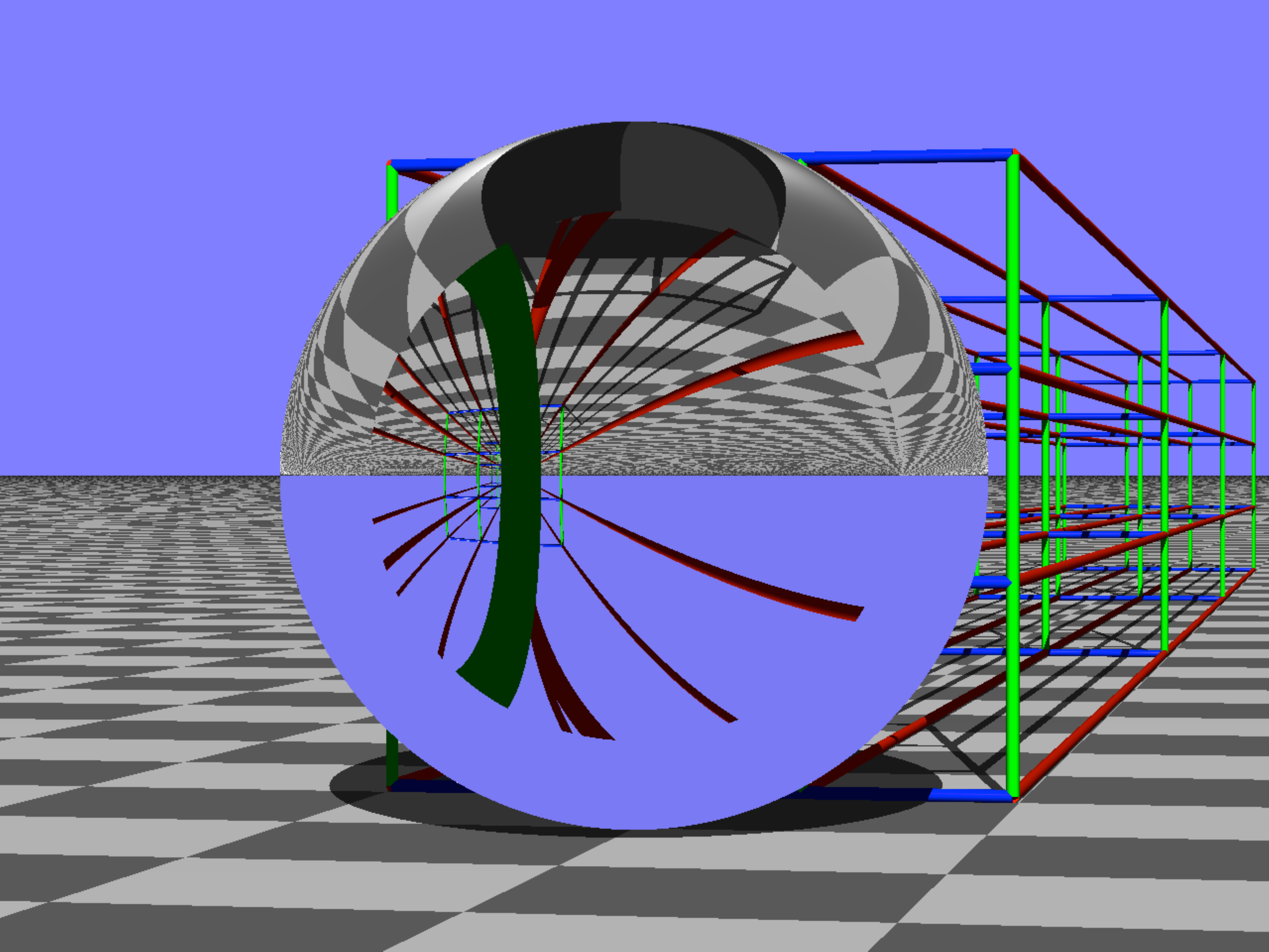}
\end{center}
\caption{\label{index-mismatch-figure}View through the lens shown in Fig.\ \ref{transparent-tunnel-figure}(b), but with the ratio between the refractive indices at the lens's surface and of the surrounding medium being $n_\mathrm{surface} / n_\mathrm{surrounding} = 1.5$ (instead of 1).
Note that the central part of the lens no longer looks like an empty tunnel.}
\end{figure}

Fig.\ \ref{index-mismatch-figure} shows an example of the view through a lens in which the surface refractive index does not match that of the surrounding medium.

\subsection{Implementation details}

\noindent
The lenses described above are implemented as classes that extend the \texttt{Sphere} class of scene objects, specifically the classes \texttt{EatonLens}, \texttt{LuneburgLens}, \texttt{MaxwellFisheyeLens}, and \texttt{GGRINLens}.
These classes are part of the package \texttt{optics.raytrace.sceneObjects}.

For use in TIM's interactive version, the above classes have been extended to be editable.
The editable versions of the above classes are called \texttt{EditableEatonLens}, \texttt{EditableLuneburgLens}, \texttt{EditableMaxwellFisheyeLens}, and \texttt{EditableGGRINLens}, and all are part of the package \texttt{optics.raytrace.GUI.sceneObjects}.

The calculation of the refraction actually happens in classes that describe surface properties, specifically extensions of the class \texttt{SurfacePropertyPrimitive}.
The surface-property classes that describe the above lenses are called \texttt{EatonLensSurface}, \texttt{LuneburgLensSurface}, \texttt{MaxwellFisheyeLensSurface}, and \texttt{GGRINLensSurface}.
All are part of the \texttt{optics.raytrace.surfaces} package.

\section{\label{metric-interface-section}Refraction at metric interfaces}

\subsection{\label{metric-interface-refraction-section}Metrics, Fermat's principle, and refraction at metric interfaces}

\begin{figure}
\begin{center}
\includegraphics{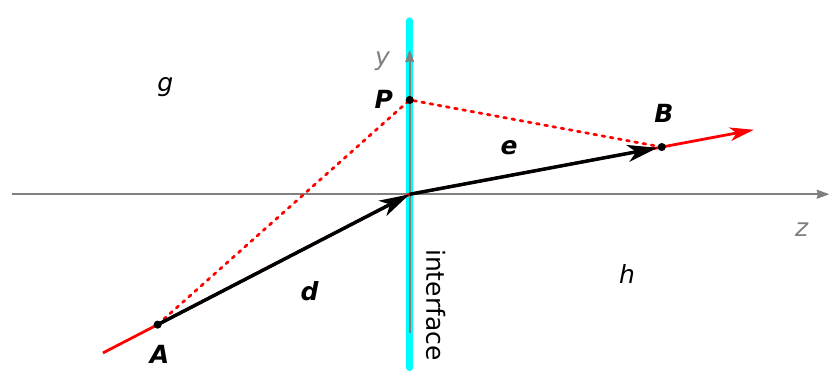}
\end{center}
\caption{\label{refraction-geometry-figure}Geometry of refraction at a metric-tensor interface.
The coordinate system is chosen such that the interface is in the $z=0$ plane and the incident light ray intersects the interface at the origin.
The diagram is drawn for positive refraction.}
\end{figure}

\noindent
In transformation optics \cite{Leonhardt-2006,Pendry-et-al-2006}, the (optical) metric of a material measures the optical path length $\rmd s$ corresponding to an infinitesimal geometrical path of length $\rmd x$, $\rmd y$ and $\rmd z$ in the $x$, $y$ and $z$ direction.
Fermat's principle demands that the optical path length along light-ray trajectories is stationary, and so creating a spatially-varying metric changes light-ray paths.
This idea has been used to design invisibility cloaks~\cite{Leonhardt-2006,Pendry-et-al-2006}.

In vacuum, the square of the optical path length is given by the Euclidean distance
\begin{equation}
\rmd s^2 = \rmd x^2 + \rmd y^2 + \rmd z^2.
\end{equation}
In transformation-optics materials, the formula for the square of the optical path length takes the more general form
\begin{equation}
\begin{aligned}
\rmd s^2
&= g_{11} \rmd x^2 + g_{22} \rmd y^2 + g_{33} \rmd z^2 \\
&+ 2 g_{12} \rmd x \rmd y + 2 g_{13} \rmd x \rmd z + 2 g_{23} \rmd y \rmd z \\
&= \bm{\rmd r}^T \cdot g \cdot \bm{\rmd r},
\label{metric-equation}
\end{aligned}
\end{equation}
where $\bm{\rmd r} = (\rmd x, \rmd y, \rmd z)^T$ and $g$ is the (symmetric) metric tensor
% The numbers $g_{ij}$ are the elements of the (symmetric) metric tensor
\begin{equation}
g = \left( \begin{array}{ccc}
g_{11} & g_{12} & g_{13} \\
g_{12} & g_{22} & g_{23} \\
g_{13} & g_{23} & g_{33}
\end{array} \right).
\end{equation}

We stress that Eqn (\ref{metric-equation}) specifies the \emph{square} of the optical path length.
To specify the optical path length itself therefore additionally requires knowledge of the sign of the optical path length; in a material with a negative refractive index, for example, it is negative~\cite{Pendry-2000}.

We have added to TIM the ability to calculate the direction of a light ray after transmission through the interface between materials with different metrics.
There are well-established methods to calculate this light-ray direction change \cite{Schurig-et-al-2006a}.
TIM calculates the refracted light-ray direction directly from Fermat's principle.
The coordinate system is placed such that the interface lies in the $z=0$ plane and the incident light ray intersects the interface at the origin (Fig.\ \ref{refraction-geometry-figure}).
The half-space with $z < 0$ is described by the metric tensor $g$ (with elements $g_{ij}$), the metric tensor for $z > 0$ is $h$ (elements $h_{ij}$), and the incident light-ray direction is $\bm{d} = (d_x, d_y, d_z)^T$.
The aim is to calculate the outgoing light-ray direction, which we write in the form $\bm{e} = (e_x, e_y, e_z)^T$.

Fermat's principle states that the optical path length between any two points on the light-ray trajectory is stationary.
% We pick two points on either side of the interface, $\bm{A}$ in front of the interface, $\bm{B}$ behind it.
We pick the two points $\bm{A} = - \bm{d}$ and $\bm{B} = \bm{e}$, which lie on either side of the interface.
The optical path length from $\bm{A}$ to $\bm{B}$ via a point $\bm{P} = (x, y, 0)^T$
%\begin{equation}
%\bm{P} = \left( \begin{array}{c} x \\ y \\ 0 \end{array} \right)
%\end{equation}
on the interface is then $s = s_1 + s_2$, where $s_1 = \pm \sqrt{(\bm{AP})^T \cdot g \cdot \bm{AP}}$ (where $\bm{AP} = \bm{P} - \bm{A}$) is the optical path length from $\bm{A}$ to $\bm{P}$, $s_2 = \pm \sqrt{(\bm{PB})^T \cdot h \cdot \bm{PB}}$ (where $\bm{PB} = \bm{B} - \bm{P}$) is the optical path length from $\bm{P}$ to $\bm{B}$.
The signs of $s_1$ and $s_2$ indicate whether or not these optical path lengths are positive or negative, of course.
Fermat's principle states that, at $(x, y) = (0, 0)$ (i.e.\ at the origin, where the actual light ray intersects the interface),
\begin{align}
\frac{\partial \sqrt{(\bm{AP})^T \cdot g \cdot \bm{AP}}}{\partial x}
\pm \frac{\partial \sqrt{(\bm{PB})^T \cdot h \cdot \bm{PB}}}{\partial x}
&= 0, \label{Fermat-x-equation-1} \\
\frac{\partial \sqrt{(\bm{AP})^T \cdot g \cdot \bm{AP}}}{\partial y}
\pm \frac{\partial \sqrt{(\bm{PB})^T \cdot h \cdot \bm{PB}}}{\partial y}
&= 0, \label{Fermat-y-equation-1}
%\frac{\partial s_1}{\partial x} + \frac{\partial s_2}{\partial x} = 0, \quad
%\frac{\partial s_1}{\partial y} + \frac{\partial s_2}{\partial y}= 0.
\end{align}
where the `$+$' signs must be chosen if $s_1$ and $s_2$ have the same signs (positive refraction), and the `$-$' signs must be chosen if $s_1$ and $s_2$ have the opposite sign (negative refraction).
Allowing negative refraction enables raytracing through many interesting interfaces, for example those representing interfaces between media with different signs of the refractive index such as a Veselago lens \cite{Veselago-1968,Pendry-2000}.
To characterise a metric interface fully, we therefore state the metric tensors on both sides and whether the refraction is positive or negative.

Evaluating the first term of Eqn (\ref{Fermat-x-equation-1}) gives
\begin{align}
\frac{\partial \sqrt{(\bm{AP})^T \cdot g \cdot \bm{AP}}}{\partial x}
= \frac{d_x g_{11} + d_y g_{12} + d_z g_{13}}{\sqrt{\bm{d}^T \cdot g \cdot \bm{d}}}.
\label{Fermat-x-equation-2a}
\end{align}
We define the number
\begin{align}
c_x 
= \pm \frac{d_x g_{11} + d_y g_{12} + d_z g_{13}}{\sqrt{| \bm{d}^T \cdot g \cdot \bm{d} |}},
\label{c_x-equation}
\end{align}
with the `$+$' sign in the case of positive refraction and the `$-$' sign for negative refraction.
Assuming, for the moment, that the term under the square root on the right-hand side of Eqn (\ref{Fermat-x-equation-2a}) is positive, $c_x$ represents the first term of Eqn (\ref{Fermat-x-equation-1}).
Note that $c_x$ depends only on the direction of the incident light ray and the material in which it travels. 
In terms of $c_x$, Eqn (\ref{Fermat-x-equation-1}) becomes
\begin{align}
c_x
&= \frac{\partial \sqrt{(\bm{PB})^T \cdot h \cdot \bm{PB}}}{\partial x}
% \frac{d_x g_{11} + d_y g_{12} + d_z g_{13}}{\sqrt{\bm{d}^T \cdot g \cdot \bm{d}}} &=
= \frac{e_x h_{11} + e_y h_{12} + e_z h_{13}}{\sqrt{\bm{e}^T \cdot h \cdot \bm{e}}}. \label{Fermat-x-equation}
%\frac{d_x g_{12} + d_y g_{22} + d_z g_{23}}{\sqrt{\bm{d}^T \cdot g \cdot \bm{d}}} &= -
%\frac{e_x h_{12} + e_y h_{22} + e_z h_{23}}{\sqrt{\bm{e}^T \cdot h \cdot \bm{e}}}. \label{Fermat-y-equation}
\end{align}
% It can be seen that a reversal of the direction of $\bm{e}$ is equivalent to choosing a different sign for the left-hand side, and so we can drop the $\pm$ signs in Eqns (\ref{Fermat-x-equation-1}) to (\ref{Fermat-y-equation}) on the understanding that, at the end of the calculation, the calculated direction $\bm{e}$ of the refracted light ray might have to be reversed to so that the incident light ray arrives from one side of the interface and the refracted light ray leaves from the opposite side.
% traverse the interface with the same sense as the direction of the incident light-ray direction,~$\bm{d}$.
Next, we normalise the vector $\bm{e}$ with respect to the metric $h$ such that
\begin{align}
\bm{e}^T \cdot h \cdot \bm{e} &= 1,
% &= h_{11} e_x^2 + h_{22} e_y^2 + h_{33} e_z^2 + 2 h_{12} e_x e_y + 2 h_{13} e_x e_z + 2 h_{23} e_y e_z \nonumber \\
% &= h_{11} e_x^2 + h_{22} e_y^2 + h_{33} e_z^2 + 2 (h_{12} e_x e_y + h_{13} e_x e_z + h_{23} e_y e_z ) \nonumber \\
\label{e-normalisation-equation}
\end{align}
so that, on the right-hand side of Eqn (\ref{Fermat-x-equation}), the denominator becomes one (thereby also ensuring that the expression under the square root is positive).
Eqn (\ref{Fermat-x-equation}) then simplifies to
\begin{equation}
c_x = e_x h_{11} + e_y h_{12} + e_z h_{13}.
\label{e-cx-equation}
\end{equation}
Similarly we get from Eqn (\ref{Fermat-y-equation-1})
\begin{equation}
c_y = e_x h_{12} + e_y h_{22} + e_z h_{23}.
\label{e-cy-equation}
\end{equation}
Together, Eqns (\ref{e-normalisation-equation}), (\ref{e-cx-equation}) and (\ref{e-cy-equation}) determine the direction $\bm{e}$ of the refracted light ray.

We solve these equations as follows.
First we solve Eqns (\ref{e-cx-equation}) and (\ref{e-cy-equation}) for $e_x$ and $e_y$ to find
\begin{align}
e_x
&= \frac{c_x h_{22} - c_y h_{12} - e_z h_{13} h_{22} + e_z h_{12} h_{23}}{h_{11} h_{22} - h_{12}^2}, \label{e-cx-equation-1} \\
e_y
&= \frac{c_y h_{11} - c_x h_{12} - e_z h_{23} h_{11} + e_z h_{12} h_{13}}{h_{11} h_{22} - h_{12}^2} \label{e-cy-equation-1}.
\end{align}
Substitution into Eqn (\ref{e-normalisation-equation}) gives, after some manipulation (which assumes that $h_{12}^2 - h_{11} h_{22} \neq 0$),
\begin{equation}
a e_z^2 + c = 0
\label{quadratic-ez-equation}
\end{equation}
where
\begin{align}
a &= h_{13}^2 h_{22} - 2 h_{12} h_{13} h_{23} + h_{12}^2 h_{33} + h_{11} h_{23}^2 - h_{11} h_{22} h_{33}, \label{a-equation} \\
c &= - c_y^2 h_{11} + 2 c_x c_y h_{12} - h_{12}^2 -c_x^2 h_{22} + h_{11} h_{22}. \label{c-equation}
\end{align}
The real solutions, which exist provided that $-c / a \geq 0$, are
\begin{equation}
e_z  = \pm \sqrt{-\frac{c}{a}}. \label{ez-equation}
\end{equation}
The correct solution is that which has the same sign as $d_z$, such that the incident ray hits one side of the interface and the refracted ray leaves from the other side.

If $-c / a < 0$, no real solution exists.
This is a generalisation of the situation when total internal reflection (TIR) occurs at refractive-index interfaces, and so in this case TIM assumes that TIR occurs.

\subsection{\label{metric-interface-refraction-imaginary-section}Imaginary optical path lengths}

\noindent
If the expression under the square root in Eqn (\ref{c_x-equation}), $\bm{d}^T \cdot g \cdot \bm{d}$, is negative (which corresponds to the metric tensor not being positive-definite, like the Lorentzian metric), it is sometimes still possible to find values for $e_x$, $e_y$ and $e_z$ which are all real-valued and which solve Eqn (\ref{Fermat-x-equation-1}).
However, this solution corresponds to imaginary optical path lengths.
It is not clear, what (if anything) it represents.
We have chosen to program this solution into TIM, but give the user the choice whether or not to allow it.
If the user does not allow it, and $\bm{d}^T \cdot g \cdot \bm{d} < 0$, TIM again assumes that TIR occurs.

After evaluating the derivatives (as above), Eqn (\ref{Fermat-x-equation-1}) becomes
\begin{equation}
\pm \frac{d_x g_{11} + d_y g_{12} + d_z g_{13}}{\sqrt{\bm{d}^T \cdot g \cdot \bm{d}}} =
\frac{e_x h_{11} + e_y h_{12} + e_z h_{13}}{\sqrt{\bm{e}^T \cdot h \cdot \bm{e}}}.
\label{Fermat-x-equation-2b}
\end{equation}
The expression under the square root on the left-hand side is negative, but if the expression under the square root on the right-hand side, $\bm{e}^T \cdot h \cdot \bm{e}$, is also negative, a common factor $\rmi$ can be cancelled and Eqn (\ref{Fermat-x-equation}) becomes
\begin{equation}
\pm \frac{d_x g_{11} + d_y g_{12} + d_z g_{13}}{\sqrt{- (\bm{d}^T \cdot g \cdot \bm{d})}} =
\frac{e_x h_{11} + e_y h_{12} + e_z h_{13}}{\sqrt{- (\bm{e}^T \cdot h \cdot \bm{e})}}.
\label{Fermat-x-alt-equation}
\end{equation}
Thanks to the absolute value under the square root in the denominator we introduced in the definition of $c_x$, Eqn (\ref{c_x-equation}), the left-hand side of Eqn (\ref{Fermat-x-alt-equation}), which depends only on the incident ray, equals $c_x$ again.
We again normalise $\bm{e}$ such that the term in the square root of the denominator on the side that depends only on the outgoing ray direction becomes one, which now requires
\begin{equation}
\bm{e}^T \cdot h \cdot \bm{e} = -1.
\label{e-alt-normalisation-equation}
\end{equation}

Eqns (\ref{e-cx-equation}) and (\ref{e-cy-equation}) then still hold, and so $e_x$ and $e_y$ can still be calculated from Eqns (\ref{e-cx-equation-1}) and (\ref{e-cy-equation-1}), and the normalisation equation can again be written in the form of Eqn (\ref{ez-equation}), but with a slightly different coefficient $c$ in which two terms have changed sign (compared to Eqn (\ref{c-equation})):
\begin{align}
c &= - c_y^2 h_{11} + 2 c_x c_y h_{12} + h_{12}^2 - c_x^2 h_{22} - h_{11} h_{22}.
\end{align}

\subsection{General coordinates}

%\textbf{TIM calculates refraction at metric interfaces for light which is travelling in the positive $z$-direction. It is possible to calculate refraction for metric interfaces which do not lie orthogonal to the $z$-axis. If for example the interface is rotated by an angle $\theta$ around the $x$-axis then the metric, $g$, can be transformed as follows
%\begin{equation}\label{metric-transform-equation}
%g'={\Lambda_{x}^{T}(\theta)}g{\Lambda_{x}(\theta)},
%\end{equation}
%where $\Lambda$ is a rotation matrix for rotation around the $x$-axis. This will still produce the correct imaging and metric $h$ on the side after the interface, so long as the light ray direction of the incident light ray is also transformed for the new coordinate system as below}
%\begin{equation}
%\mathbf{d'}={\Lambda_{x}(\theta)}\mathbf{d}
%\end{equation}

\noindent
So far in this section we have assumed that the interface is in the $z=0$ plane.
This assumption greatly simplifies the calculation of the light-ray direction behind a metric interface.
We now generalise these results to an arbitrary interface.
We achieve this by defining a local surface coordinate system in which the tangent plane to the surface at the incident ray's intersection point is the equivalent of the $z=0$ plane;
calculating the incident light-ray direction and the metric tensors on both sides of the interface in this surface coordinate system;
calculating the refracted light-ray direction in the surface coordinate system (according to the calculation outlined in sections \ref{metric-interface-refraction-section} and \ref{metric-interface-refraction-imaginary-section});
and finally transforming the refracted light-ray direction into the global $(x, y, z)$ coordinate system.

We call the unit base vectors of the local coordinate system $\hat{\bm{u}}$, $\hat{\bm{v}}$ and $\hat{\bm{w}}$, and choose $\hat{\bm{w}}$ to be perpendicular to the surface at the intersection point.
The local tangent plane is therefore the $w=0$ plane, which from now on takes over the role played by the $z=0$ plane in sections \ref{metric-interface-refraction-section} and \ref{metric-interface-refraction-imaginary-section}.
In TIM, such a coordinate system can easily be calculated for each point on a surface (see Fig.\ 7 in Ref.\ \cite{Lambert-et-al-2012}).

The calculation of light-ray-direction vectors between the bases is standard vector maths.
In terms of the metric tensor $g$ in the $(x, y, z)$ basis, the metric tensors in the $(u, v, w)$ basis, $g^\prime$, is
\begin{equation}
g^\prime = \Lambda^T g \Lambda,
\label{general-metric-transform-equation}
\end{equation}
where
\begin{equation}
\Lambda
= \begin{pmatrix} \hat{\bm{u}} & \hat{\bm{v}} & \hat{\bm{w}} \end{pmatrix}
= \begin{pmatrix}
\partial x / \partial u & \partial x / \partial v &\partial x / \partial w  \\
 \partial y / \partial u & \partial y / \partial v &\partial y / \partial w  \\
 \partial z / \partial u & \partial z / \partial v &\partial z / \partial w
\end{pmatrix}
\end{equation}
is the Jacobian matrix.

\subsection{Implementation in TIM}

\noindent
In TIM, metric-tensor interfaces are represented by an object of class \texttt{MetricTensorInterface}.
The calculation of the refracted light-ray direction $\bm{e}$ happens in the \texttt{getRefractedRayDirection} method.
Note that the user can choose whether or not the metric tensors on both sides of the interface are given in the global coordinate system or in the surface coordinate system.
The latter is useful for defining homogeneous, but non-planar, surfaces.

\begin{figure}
\begin{center}
\begin{tabular}{cc}
\raisebox{4.5cm}{(a)} & \includegraphics[width=0.8 \columnwidth]{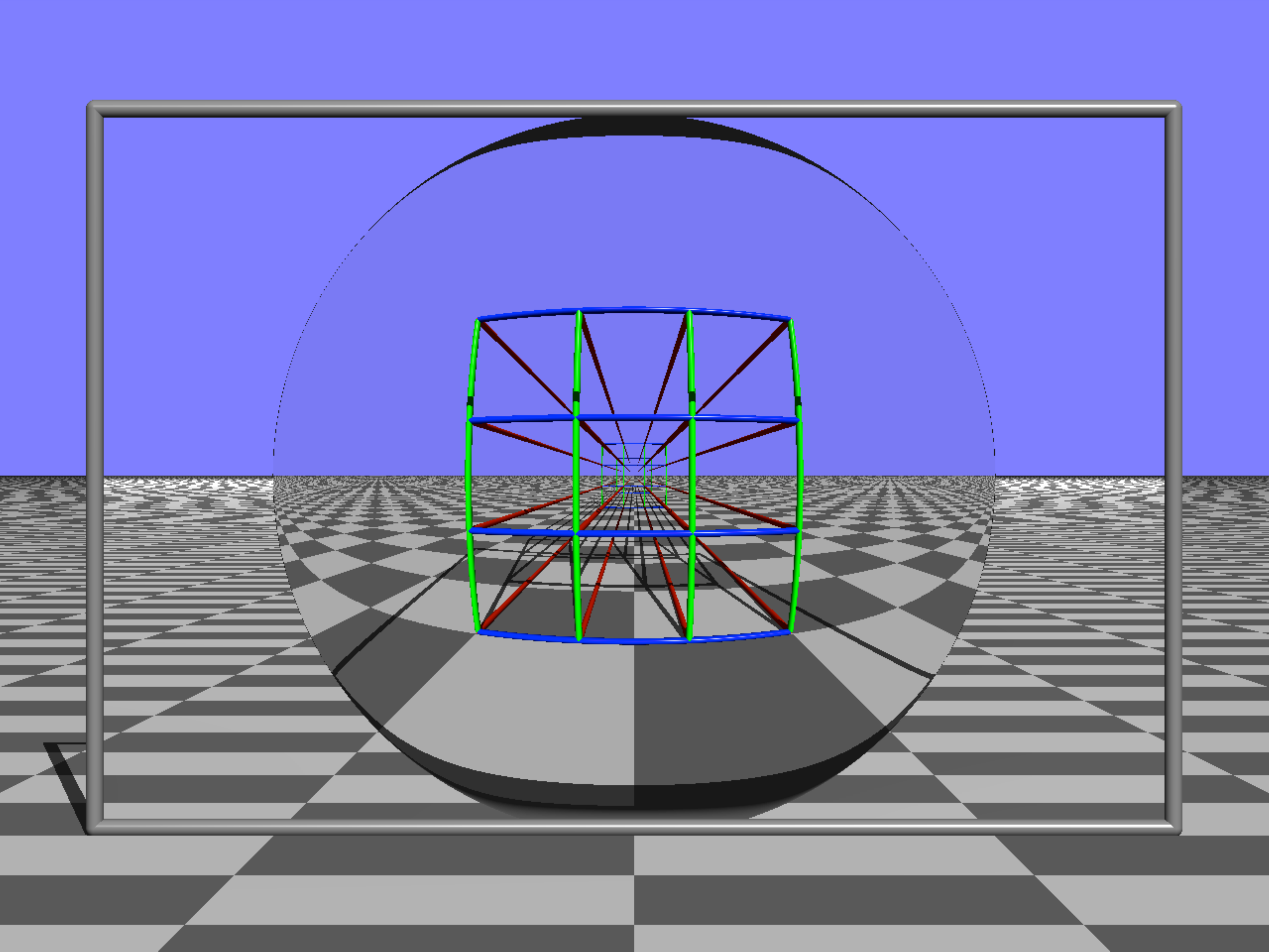} \\
\raisebox{4.5cm}{(b)} & \includegraphics[width=0.8 \columnwidth]{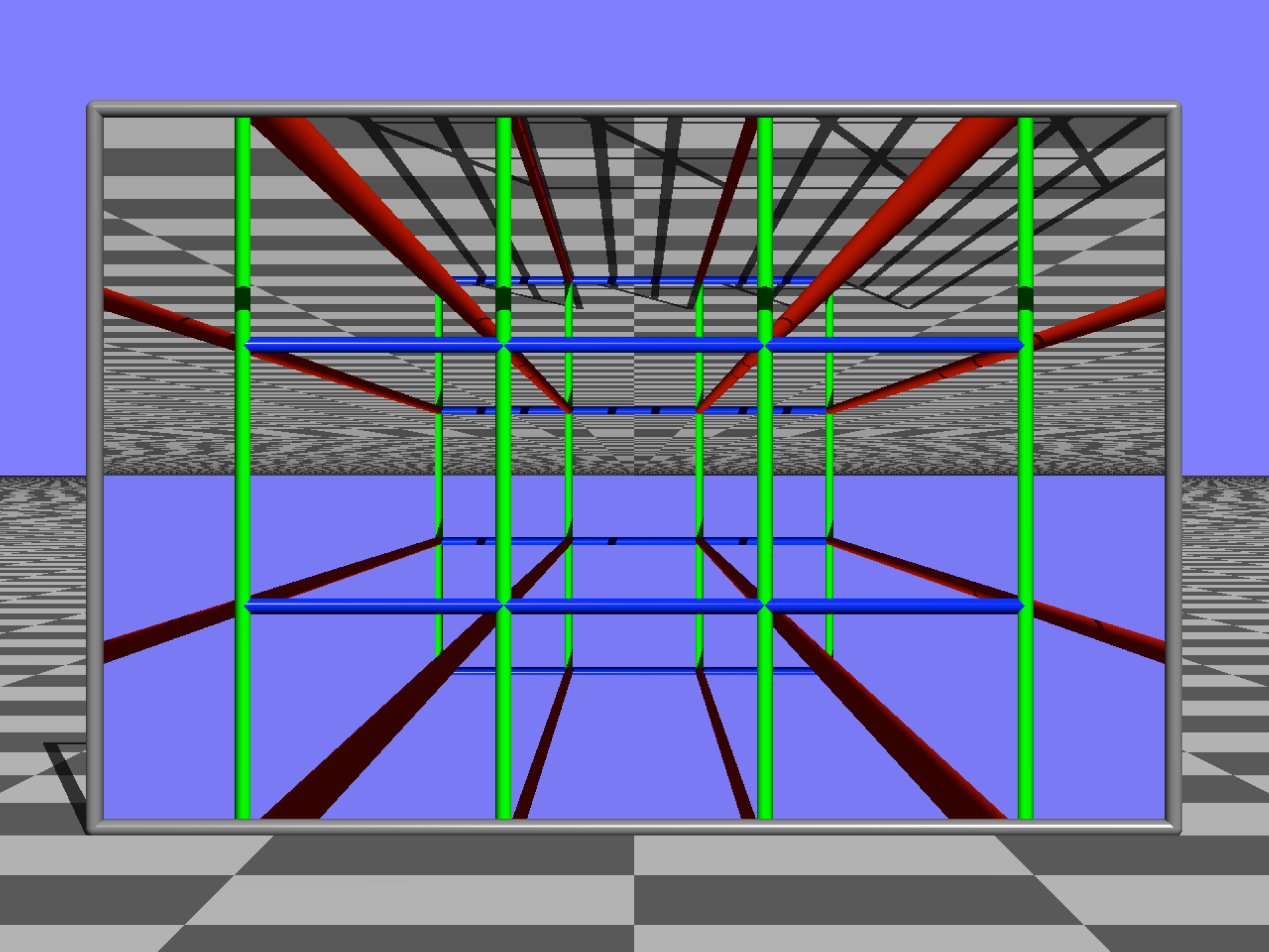} \\
\raisebox{4.5cm}{(c)} & \includegraphics[width=0.8 \columnwidth]{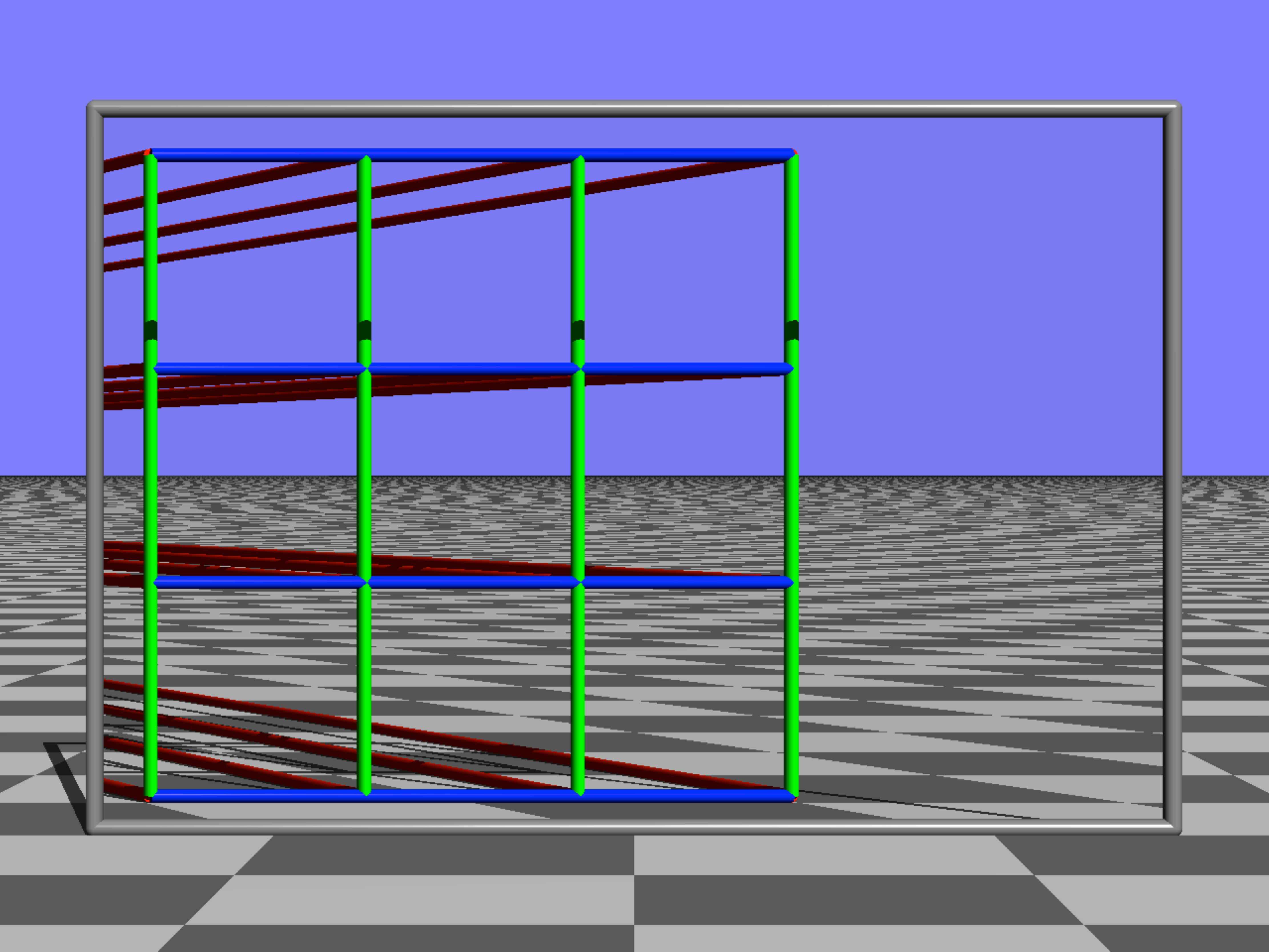}
\end{tabular}
\end{center}
\caption{\label{metric-tensor-interface-figure}Simulated view through different metric interfaces.
(a)~Metric interface equivalent to refractive index $n = 10$ in front of the interface and $n = 1$ behind it.
Total internal reflection (TIR) is clearly visible outside the central disc.
(b)~Metric interface representing a Veselago lens \cite{Veselago-1968}, a refractive-index interface between $n = -1$ in front and $n = +1$.
(c)~Interface between a space sheared by $45^\circ$ in the $(x, z)$ projection in front and a Euclidean space behind.}
\end{figure}

Fig.\ \ref{metric-tensor-interface-figure} shows a few examples of the simulated view through various metric-tensor interfaces.
The procedure to create these (and similar) views is as follows.
Start up TIM and click on the ``Edit scene'' button.
In the ``Initialise scene to...'' drop-down menu select ``METATOY science (lattice behind METATOY window)''.
The list of scene objects now contains an object called ``METATOY window''; double-click on this object to edit its parameters.
The window is placed such that it fills a good portion of the default camera.
By default, the window's surface type is ``Rotating'' (which simulates light-ray rotation around the window normal \cite{Hamilton-et-al-2009}); click on the surface-type drop-down menu and select ``Metric-tensor interface''.
The elements of the metric tensor ``inside'' and ``outside'' the window\footnote{In TIM, every surface has an inside and an outside, defined by the surface normal which, by definition, points outwards.
From the default camera position, the default window is seen from the inside.
The direction of the surface normal can be visualised by right-clicking on a surface in the rendered image and selecting, in the pop-up menu that appears, ``Add surface-coordinate axes''.
Rendering the scene again now also shows three arrows pointing along the three coordinate axes defining the surface coordinate system; the outwards surface normal is the blue arrow.} can then be inspected and edited in the tabbed pane.

\begin{figure}
\begin{center}
\includegraphics[width=\columnwidth]{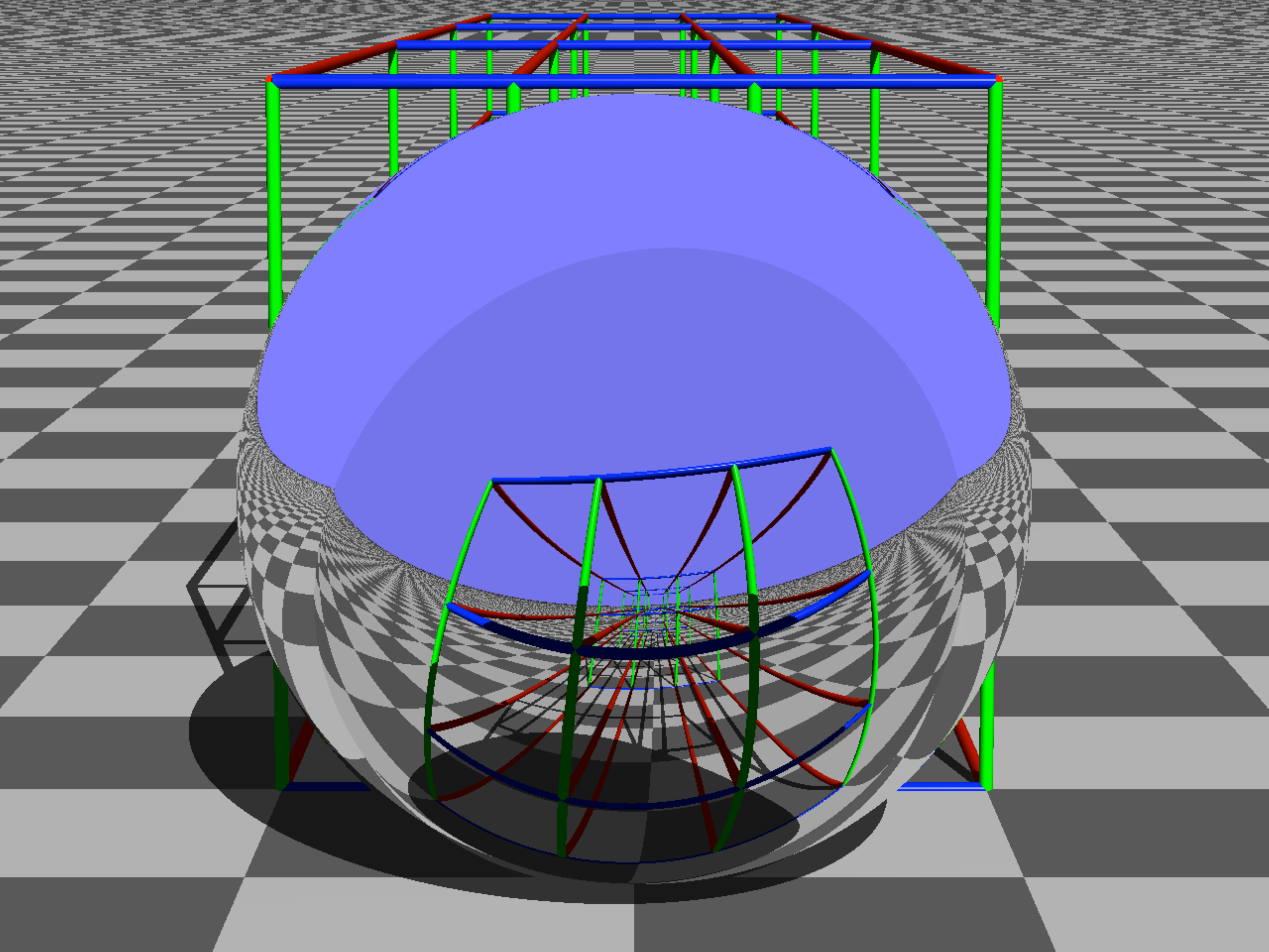} \\
\includegraphics[width=\columnwidth]{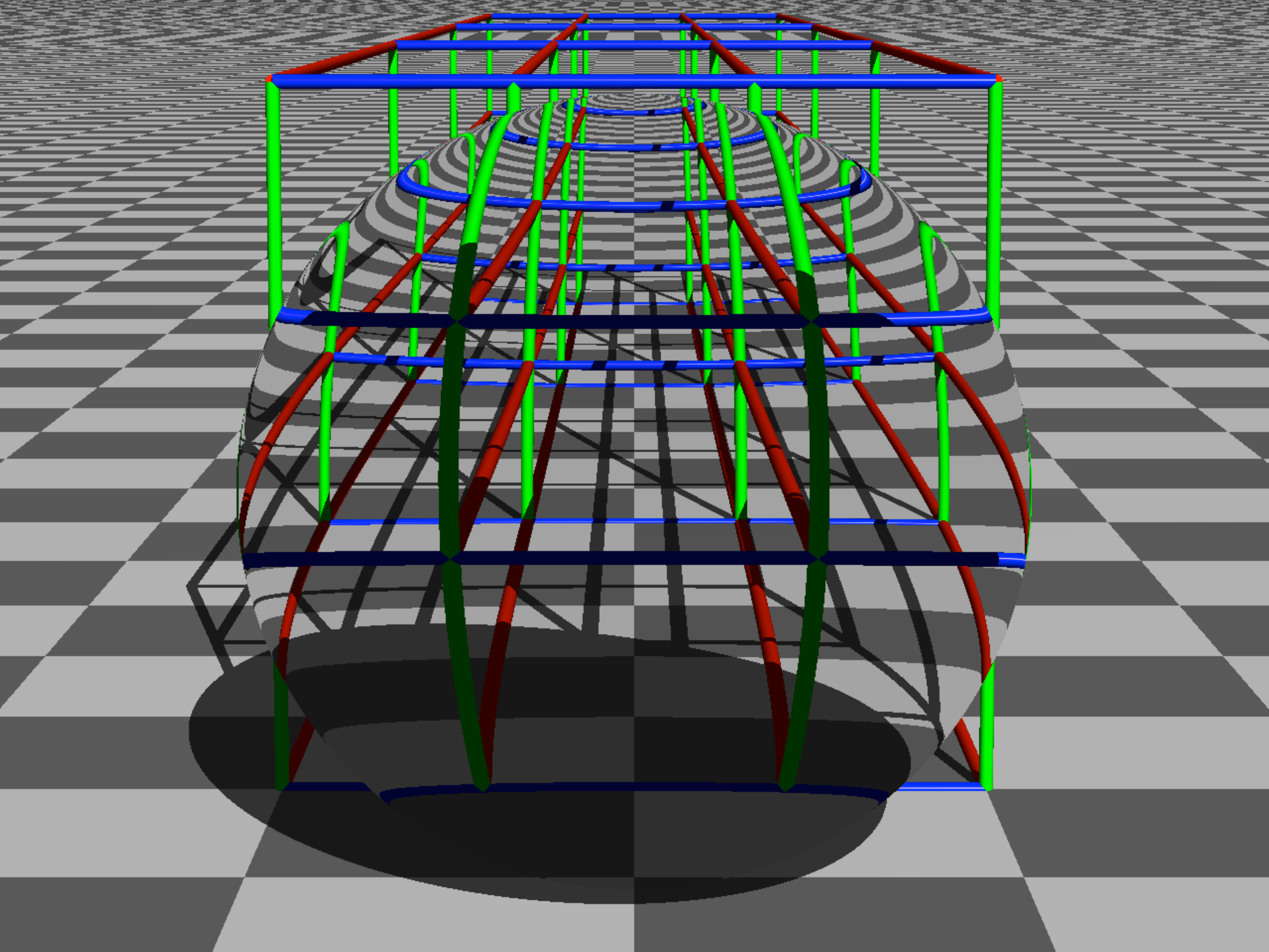}
\end{center}
\caption{\label{space-bubble-figure}View of TIM's standard cylinder lattice through ``bubbles'' of space with a non-Euclidean metric.
(Top)~The elements of the metric tensor inside the sphere, expressed in the global coordinate basis, are $g_{11} = 1$, $g_{12} = g_{21} = 0.2$, $g_{13} = g_{31} = 0$, $g_{22} = 1.2$, $g_{23} = g_{32} = -0.5$, $g_{33} = 0.5$.
TIR is visible near the edge of the sphere.
(Bottom)~The inside of the sphere is described by the Minkowski-like metric $\mathrm{diag}(-1, 1, 1)$.}
% sphere centred at (0, 0, 9), metric on inside ((1, 0.2, 0), (0.2, 1.2, -0.5), (0, -0.5, 0.5))
% standard lattice
% aperture centre (0, 1.3, 0), view direction (0, -0.14, 1), anti-aliasing quality good
\end{figure}

% can also be applied to curved surfaces.
Fig.\ \ref{space-bubble-figure} illustrates

\section{\label{gCLAs-section}Refraction with generalised confocal lenslet arrays}

\noindent
Another extension we recently added to TIM is the capability to simulate the transmission of light rays through generalised confocal lenslet arrays (gCLAs) \cite{Hamilton-Courtial-2009b}.
This extension has already been discussed elsewhere \cite{Oxburgh-Courtial-2013b}, and so, after a brief summary of gCLAs and their effect on light rays, we concentrate on details not already discussed.

\subsection{Generalised confocal lenslet arrays (gCLAs)}

\noindent
Telescopes have the property that light rays that are parallel when incident are parallel again when leaving.
In other words, the direction change does not depend on the precise point where a ray hits the telescope.
There is a ray offset, but it is small when the telescopes themselves are small, and so the effect on light rays of a planar array of small telescopes (``telescopelets'') with their optical axes perpendicular to the plane is essentially generalised refraction, and so such components are examples of METATOYs \cite{Hamilton-Courtial-2009}.
Perhaps the easiest way to build such arrays of telescopelets is from two arrays of lenslets that share a common focal plane, or confocal lenslet arrays (CLAs) \cite{Courtial-2008a}.
CLAs have interesting imaging properties \cite{Courtial-2008a,Courtial-2009b}, and they have been demonstrated experimentally \cite{Courtial-et-al-2010}.

\begin{figure}
\begin{center}
\includegraphics{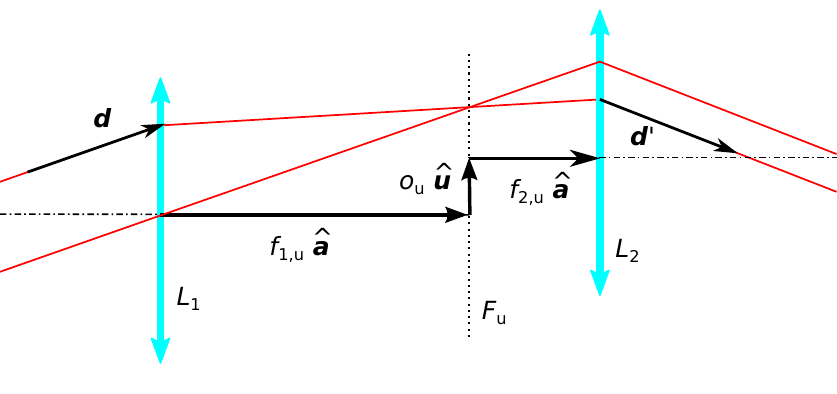}
\end{center}
\caption{\label{gCLAs-figure}Ray propagation through one of the telescopelets in generalised confocal lenslet arrays (gCLAs).
The diagram is an orthographic projection into the plane spanned by the vectors $\hat{\bm{a}}$ and $\hat{\bm{u}}$, a unit vector in the $u$ direction, one of the two directions perpendicular to $\hat{\bm{a}}$ (the other being the $v$ direction).
The telescopelet is formed by two lenses, $L_1$ and $L_2$, whose optical axes (dash-dotted horizontal lines) are parallel (the unit vector $\hat{\bm{a}}$ points in the direction of the optical axes), but offset relative to each other, by a distance $o_u$ in the $u$ direction and a distance $o_v$ in the $v$ direction.
In general, the lenses have different focal lengths in the $u$ and $v$ directions, but the focal lengths of the two lenses in the $u$ direction add up to the separation between the lenses, as do the focal lengths in the $v$ direction.
$F_u$ is the common focal plane in the $(u, a)$ projection.
The solid red lines show two light-ray trajectories, both incident with direction $\bm{d}$ and leaving with direction $\bm{d}^\prime$.}
\end{figure}

The telescopelets in CLAs can be generalised in a number of ways while retaining their basic METATOY character (of introducing a light-ray-direction change that is independent of the precise position where a light ray enters the METATOY while introducing a small offset) \cite{Hamilton-Courtial-2009b}.
The generalisations are as follows (see Fig.\ \ref{gCLAs-figure}):
\begin{enumerate}
\item Each lens can be replaced by a combination of cylindrical lenses with their cylinder axes in two transverse directions, $u$ and $v$, and so each lens has two focal lengths.
The focal lengths in the $u$ direction of both lenslets have to add up to the separation between the lenslets, and so do the focal lengths in the $v$ direction.
\item The lenses' optical axes can be offset relative to each other.
\item The entire telescope can be rotated.
\end{enumerate}

In Ref.\ \cite{Hamilton-Courtial-2009b}, the telescopes were considered to be aligned initially such that their optical axis points in the $z$ direction, and so that the cylinder axes are aligned in the $x$ and $y$ directions.
The parameters necessary to describe a general telescopelet aligned in this way are
$\eta_x$, the ratio of the focal lengths in the $x$ direction of the second and first lens, multiplied by $(-1)$;
$\eta_y$, the ratio of the focal lengths in the $y$ direction of the second and 1st lens, again multiplied by $(-1)$;
$\delta_x$, the $x$ offset of the optical axis of the second lens relative to that of the first, divided by the first lens's focal length in the $x$ direction;
and $\delta_y$, the $y$ offset of the optical axis of the second lens relative to that of the first, divided by the first lens's focal length in the $y$ direction.
A general rotation, described by Euler angles $\phi$, $\theta$ and $\psi$, makes such a telescopelet completely general.
The seven parameters described above are, of course, the seven degrees of freedom that make gCLAs such versatile components.

In Ref.\ \cite{Oxburgh-Courtial-2013b}, gCLAs were described slightly differently, namely by
$\hat{\bm{a}}$, a unit vector in the direction of the optical axis \emph{after} rotation;
$\hat{\bm{u}}$, a unit vector in the direction of the $x$ direction after rotation;
$\eta_u$, the focal-length ratio in the $u$ direction of the second and first lenses, multiplied by $(-1)$, and $\eta_v$, the equivalent ratio for the $v$ direction, which is perpendicular to both the $a$ and $u$ direction (a unit vector in the $v$ direction is defined as $\hat{\bm{v}} = \hat{\bm{a}} \times \hat{\bm{u}}$);
and by $\delta_u$, the $u$ offset of the second optical axis relative to the first, divided by the focal length in the $u$ direction of the first lens, and $\delta_v$, the equivalent ratio for the $v$ direction.
We use this latter description, for which the law of refraction is \cite{Oxburgh-Courtial-2013b}
\begin{align}
\bm{d}^\prime
&= \frac{(\bm{d} \cdot \hat{\bm{u}}) / (\bm{d} \cdot \hat{\bm{a}}) - \delta_u}{\eta_u} \hat{\bm{u}}
+ \frac{(\bm{d} \cdot \hat{\bm{v}}) / (\bm{d} \cdot \hat{\bm{a}}) - \delta_v}{\eta_v} \hat{\bm{v}}
+ \hat{\bm{a}}.
\label{vector-form-generalised-law-of-refraction}
\end{align}

\subsection{Implementation details}

\noindent
In TIM, generalised confocal lenslet arrays are represented by the class \texttt{ConfocalLensletArrays}, which is a non-abstract implementation of the abstract \texttt{SurfaceProperty} class.
The code calculating the refraction is part of the \texttt{getColour} method, which handles passage of an incident light ray through the surface.
The calculation of the direction of the refracted ray according to Eqn (\ref{vector-form-generalised-law-of-refraction}) is mostly straightforward.
The few subtleties are perhaps worth mentioning.

Firstly, TIM does not ask the user to enter the normalised, and perpendicular, vectors $\hat{\bm{a}}$ and $\hat{\bm{u}}$.
Instead, TIM asks for vectors $\bm{a}$ and $\bm{u}$, which are related to $\hat{\bm{a}}$ and $\hat{\bm{u}}$ as follows.
The unit vector $\hat{\bm{a}}$ is simply the vector $\bm{a}$, normalised.
In contrast, $\hat{\bm{u}}$ is not generally a normalised version of $\bm{u}$:
TIM calculates $\hat{\bm{u}}$ by projecting $\bm{u}$ into a plane perpendicular to $\hat{\bm{a}}$, and then normalising it.

%actually the direction can be $\pm$ the direction the vectors $\hat{\bm{x}}$, $\hat{\bm{y}}$ and $\hat{\bm{z}}$ would point in after Euler rotation characterised by the Euler angles $(\varphi, \theta, \psi)$.
%This sign does not matter for the $\eta$s, but it does matter for the $\delta$s.
%not necessarily right-handed coordinate system

Secondly, TIM gives the user the choice to define the vectors $\bm{a}$ and $\bm{u}$ in different bases, namely either in terms of the ``global'' ($x$, $y$, $z$) coordinate system or in terms of the ``local'' surface coordinates (two directions tangential to the surface and a surface normal; see Fig.\ 7 in Ref.\ \cite{Lambert-et-al-2012}).
The latter allows modelling of a non-planar, but homogeneous, gCLA surface.

Finally, it is important in which order a ray encounters the two lenslet arrays that make up the gCLAs.
Unlike other surfaces, this order is not decided by whether or not the light ray reaches the surfaces from the inside or the outside, but instead by the sense of the vector $\bm{a}$, which TIM interprets as pointing from lenslet array 1 to lenslet array 2 (see Fig.\ \ref{gCLAs-figure}).
TIM establishes whether or not $\bm{a}$'s and $\bm{d}$'s components perpendicular to the surface have the same sign; if they do, the ray first intersects lenslet array 1, otherwise lenslet array 2.
It can be shown that, in the latter case, the law of refraction is still given by Eqn (\ref{vector-form-generalised-law-of-refraction}), but with $\hat{\bm{a}}$ replaced by $-\hat{\bm{a}}$, $\eta_u$ replaced by $1/\eta_u$ (and $\eta_v$ by $1/\eta_v$), $\delta_u$ replaced by $\delta_u / \eta_u$ (and $\delta_v$ by $\delta_v / \eta_v$).

%$\eta_u$ and $\eta_v$:
%\begin{equation}
%\eta = - \frac{f_2}{f_1}
%\end{equation}
%
%$\delta_u$ and $\delta_v$:
%\begin{equation}
%\delta = \frac{o}{f_1}
%\end{equation}
%
%\begin{equation}
%\eta^\prime = - \frac{f_1}{f_2} = \frac{1}{\eta}
%\end{equation}
%
%\begin{equation}
%\delta^\prime = - \frac{o}{f_2} = \frac{o}{f_1} \left( - \frac{f_1}{f_2} \right) = \delta / \eta
%\end{equation}

\section{\label{odds-and-sods-section}Other extensions}

\noindent
We have not discussed a number of other changes we made to TIM since the publication of Ref.\  \cite{Lambert-et-al-2012}, including a number of extensions (such as the capability to model arrays of objects;
phase-conjugating surfaces;
surfaces that change the direction of light rays like a Lorentz transform;
surfaces which, when spherical, look from the outside like an Eaton lens or a Luneburg lens;
imperfect teleporting surfaces;
surfaces that refract like phase holograms;
the capability to create 3D images in various HDMI 1.4a standard frame-packing formats (i.e.\ a single image file simply consisting of the left-eye and right-eye images placed next to each other or on top of each other) compatible with 3D TVs;
the capability to move the camera, and vary its field of view, in the interactive version also;
and to switch shadow throwing on or off).
Most of these changes are minor;
others are important, and we intend to describe them in detail elsewhere.

\section{\label{conclusions-section}Conclusions}

\noindent
TIM is turning out to be a very useful tool for our research.
We are greatly enjoying extending TIM, and we are already planning further extensions.

As before, we aim to encourage others to use and extend TIM.
We hope that TIM's new capabilities will help us to achieve this aim.

% Sometimes, extensions in combination with TIM's features allow new things, such as relativistic raytracing with a camera with finite aperture.

\section{Acknowledgments}
% \ack
\noindent
% Many thanks to Martin \v{S}arbort for making available to us his submitted paper.
SO acknowledges funding from the UK's Engineering \& Physical Sciences Research Council (EPSRC).
TT acknowledges support of the grant P201/12/G028 of the Grant agency of 
the Czech Republic and of the QUEST programme grant of the EPSRC.
% This research received no specific grant from any funding agency in the public, commercial or not-for-profit sectors.

% \section*{References}

% \bibliography{/Users/johannes/Documents/work/library/Johannes}

\end{document}